\documentclass[
twocolumn, 
prb, 
reprint, 
amsfonts, 
amsmath, 
amssymb, 
superscriptaddress, 
longbibliography, 
floatfix, 
aps]{revtex4-2}
\usepackage[utf8]{inputenc} % directly input accented characters
\usepackage[T1]{fontenc} % font encoding of output, correctly hyphenates accented characters
\usepackage{lmodern} % Latin modern font 
\usepackage{graphicx}
\usepackage[dvipsnames]{xcolor}
\usepackage[colorlinks
,citecolor= magenta
,linkcolor=blue
,filecolor=blue
,urlcolor= blue
,breaklinks= true]{hyperref}
\usepackage{mathtools}
\usepackage{enumitem}
\usepackage{microtype}
\usepackage{physics}
\usepackage{esvect}
\usepackage{dutchcal}
\usepackage{booktabs}
\usepackage{cancel}
\usepackage{comment}

\usepackage{orcidlink}

\DeclareMathOperator{\ldiv}{div}
\DeclareMathOperator{\lcurl}{curl}
\DeclareMathOperator{\lgrad}{grad}
\DeclareMathOperator{\ii}{\mathbf{i}}
\newcommand{\ve}[1]{\boldsymbol{\mathrm{#1}}}
\newcommand{\m}{\text{m}}
\newcommand{\gbE}{g_{bE}}
\newcommand{\geB}{g_{eB}}

\newcommand{\physE}{\ve{E}}
\newcommand{\physB}{\ve{B}}
\newcommand{\eme}{\ve{e}}
\newcommand{\emg}{\ve{g}}
\newcommand{\ema}{\ve{a}}
\newcommand{\rr}{\mathsf r}
\newcommand{\rrp}{\mathsf r'}
\newcommand{\cT}{\mathcal T}
\newcommand{\Msym}{M_{x\bar y}}
\newcommand{\Mmat}{\mathsf M}
\newcommand{\Id}{\mathsf I}

\newcommand{\sectn}[1]{\textit{\color{blue}{#1.}}}
\newcommand{\lagr}{L}
\newcommand{\kB}{k_{B}}

\newcommand{\beginsupplement}{%
  \setcounter{equation}{0}
  \renewcommand{\theequation}{S\arabic{equation}}%
  \setcounter{table}{0}
  \renewcommand{\thetable}{S\arabic{table}}%
  \setcounter{figure}{0}
  \renewcommand{\thefigure}{S\arabic{figure}}%
  \setcounter{section}{0}
  \renewcommand{\thesection}{S-\Roman{section}}%
  \setcounter{subsection}{0}
  \renewcommand{\thesubsection}{S-\Roman{section}.\Alph{subsection}}%
  \setcounter{subsection}{0}
  \renewcommand{\thesubsubsection}{S-\Roman{section}.\Alph{subsection}.\arabic{subsubsection}}%
  \renewcommand{\refname}{Supplementary References}%
  \renewcommand{\labelenumi}{(\roman{enumi})}
  \renewcommand{\labelenumii}{(\roman{enumi}-\arabic{enumii})}
}

\graphicspath{{./Figs/}}

\begin{document}

\title{The monopole plasma resonance: a smoking gun of 3D $U(1)$ quantum spin liquids}

\author{Anish Koley\,\orcidlink{0009-0009-8766-1429}}
\email{anish.koley@uni-leipzig.de}
%\thanks{These authors contributed equally to this work.}
\affiliation{Institut f{\"u}r Theoretische Physik, Universit{\"a}t Leipzig, Br{\"u}derstra{\ss}e 16, 04103, Leipzig, Germany}

\author{Saranyo Moitra\,\orcidlink{0000-0001-7912-1961}}
\email{saranyo.moitra@uni-leipzig.de}
%\thanks{These authors contributed equally to this work.}
\affiliation{Institut f{\"u}r Theoretische Physik, Universit{\"a}t Leipzig, Br{\"u}derstra{\ss}e 16, 04103, Leipzig, Germany}

\author{Inti \surname{Sodemann Villadiego}\,\orcidlink{0000-0002-1824-5167}}
\email{sodemann@uni-leipzig.de}
\affiliation{Institut f{\"u}r Theoretische Physik, Universit{\"a}t Leipzig, Br{\"u}derstra{\ss}e 16, 04103, Leipzig, Germany}

\date{\today}

\begin{abstract}
Certain 3D $U(1)$ spin liquids, such as those arising in dipolar quantum spin ice, have an emergent monopole  which is the source of an emergent magnetic field that transforms under symmetries like an electric polarization. As a consequence, these monopoles carry a physical electric charge in their cores and form a plasma at low temperatures. Due to the monopole coupling to emergent gauge fields, the full system behaves as an electrical insulator for DC transport, but can display a sharp plasma resonance analogous to a metal at very low frequencies. This can serve as a clear fingerprint to detect these states in materials. We discuss the optimal conditions to observe this phenomenon in the 3D $U(1)$ spin-liquid candidate Ce$_2$Zr$_2$O$_7$. 
\end{abstract}

\maketitle

\sectn{Introduction} Three-dimensional $U(1)$ quantum spin liquids (QSL) are states of matter, where quantum electrodynamics fully equipped with photons, charges and monopoles, emerges from systems of coupled spins 
\cite{Wen_PRL01_OriginGaugeBosons,
Hermele.Balents_PRB04_PyrochlorePhotons$U1$, Gingras.McClarty_RPP14_QuantumSpinIce}
or other simple underlying degrees of freedom 
\cite{Motrunich.Senthil_PRL02_ExoticOrderSimple,Senthil.Motrunich_PRB02_MicroscopicModelsFractionalized,Wen_PRB03_ArtificialLightQuantum,banerjee_unusual_2008}. One of the striking features of these states is that they have an {\it absolutely robust gaplessness} beyond the symmetry breaking paradigm \cite{Wen_PRL01_OriginGaugeBosons}. For example, in conventionally ordered magnetic states the gaplessness of spin-waves is only protected by exact continuum symmetries. But since these symmetries are never exact in materials, their spin-waves are never exactly gapless. In contrast, a 3D $U(1)$ QSL features a collective excitation of spins, the photon, which cannot be gapped out by any coupling added to the Hamiltonian regardless of its symmetry \footnote{Provided that it is local (e.g. only couples a finite number of neighboring spins) and it is not too strong so as to induce a phase transition into another state of matter.}. 

%While the key ideas for the emergence of such states in condensed-matter-style lattice models go back to the pioneering works in lattice gauge theories [XXX], an important work that highlighted the promise of quantum spin ice blah blah blah.

Another amazing emergent particle in 3D $U(1)$ QSL are magnetic monopoles 
\footnote{We are using the terminology of Ref.\cite{Hermele.Balents_PRB04_PyrochlorePhotons$U1$}, where the spin-ice rule violations would be viewed the emergent electric charges (i.e. not the magnetic monopoles). The magnetic monopoles have also been referred to as visons \cite{Gingras.McClarty_RPP14_QuantumSpinIce}.}. Unlike photons, monopoles are {\it non-local} particles: they are the end-points of  tensionless Dirac strings  \footnote{Namely their energy cost approaches a constant in the limit of long strings}, 
which act as point sources of emergent  magnetic fields, which are ultimately local vectorial observables. The nature of these vectorial observables depends on the specific microscopic setting on which a $U(1)$ QSL emerges: in some cases it could behave like a magnetization, in some like an electric polarization, and in others like neither of the two
\cite{Villadiego_PRB21_PseudoscalarU1Spin}. 
Over the years, these possibilities have inspired creative investigations of the couplings and phenomena in $U(1)$ QSLs driven by physical electro-magnetic fields \cite{Ioffe.Larkin_PRB89_GaplessFermionsGaugeb,
Motrunich_PRB06_OrbitalMagneticField,
Ng.Lee_PRL07_PowerLawConductivityMott,
Bulaevskii.Khomskii_PRB08_ElectronicOrbitalCurrents,
Khomskii_NC12_ElectricDipolesMagnetic,
Potter.Lee_PRB13_MechanismsSubgapOptical,
Lantagne-Hurtubise.Moessner_PRB17_ElectricFieldControl,
Fu.Perkins_PRB17_FingerprintsQuantumSpin,
Sodemann.Senthil_PRB18_QuantumOscillationsInsulators,
Rao.Sodemann_PRB19_CyclotronResonanceMott,
Villadiego_PRB21_PseudoscalarU1Spin,
Khoo.SodemannVilladiego_NJP22_CorrigendumUniversalShear,
Khoo.Villadiego_PRB22_ProbingQuantumNoise,
Laumann.Moessner_PRB23_HybridDyonsInvertedb,
Wu.Ong_N24_ChargeNeutralElectronicExcitations}. In particular, Laumann and Moessner \cite{Laumann.Moessner_PRB23_HybridDyonsInvertedb} pointed out that in some 3D $U(1)$ QSLs the emergent magnetic field has the same symmetries as a physical electric polarization, and, as a consequence, the monopole acts as a source of electric polarization with a finite electric charge bound to its core. 

This interesting conclusion from Ref.\cite{Laumann.Moessner_PRB23_HybridDyonsInvertedb} is the key starting point of our work. Our goal is to determine the electrical conductivity of a 3D $U(1)$ QSL that harbours an itinerant charged fluid of magnetic monopoles, i.e. a {\it monopole plasma}. Such monopole fluid will be inevitably present in any 3D $U(1)$ QSL at finite temperatures via thermal activation of monopoles. Naively, one might think that such fluid of charged monopoles would render the QSL metallic, but we will see that the total conductivity remains insulating in the DC frequency limit. 
However, we have found the remarkable conclusion that the electrical conductivity of the 3D $U(1)$ QSL with charged monopoles can display a sharp metallic-like plasma resonance at a small but finite frequency estimated to be:
\begin{equation}\label{plasmafreq}
    \hbar \omega_p \simeq  2 \pi \sqrt{W_\m \mu^{-1}}  \sqrt{a^3n_\m(T) },
\end{equation} 
where $W_\m$ and $\mu^{-1}$ are respectively energy scales of the monopole bandwidth and the emergent magnetic field stiffness  \footnote{Namely the energy scale obtained from the coefficient in front of the Maxwell-like term associated with the low energy cost for emergent magnetic fields, which in QSI is controlled by the ring exchange energy scale \cite{Hermele.Balents_PRB04_PyrochlorePhotons$U1$} and is also typically comparable to the monopole gap $\Delta_\m$.}, $a$ is the lattice constant, and $n_\m(T)$ the density of thermally excited monopoles at temperature $T$. This is a remarkable conclusion because it means that monopoles form a soft emergent charged fluid in an otherwise insulating state of local pseudo-spin moments. Namely, this resonance can occur at extremely small temperature and frequency scales in comparison to the ordinary gaps for electrons in the insulating QSL material.

 To appreciate how extreme this can be, let us consider the case of Ce$_2$Zr$_2$O$_7$, which belongs to a class of cerium based pyrochlores for which there is evidence supporting the realization of a 3D $U(1)$ QSL \cite{Gaudet.Gaulin_PRL19_QuantumSpinIceb,
 Gao.Dai_NP19_ExperimentalSignaturesThreedimensionalb,
 Baidya.Gopalakrishnan_JPCB07_OxygenReleaseStoragePropertiesa,
 Sibille.Fennell_NP20_QuantumLiquidMagneticb,
 Smith.Gaulin_PRX22_Case$mathrmU1_ensuremathpi$Quantuma,
 Poree.Sibille_PRM22_CrystalfieldStatesDefectb,
 Yahne.Ross_PRX24_DipolarSpinIce,
 Smith.Gaulin_PRB23_QuantumSpinIceb,
 Gao.Dai_NP25_NeutronScatteringThermodynamica}, and which belongs to the interesting class of Dipolar-Octupolar quantum spin ice (QSI) systems \cite{Huang.Hermele_PRL14_QuantumSpinIcesb,
 Li.Chen_PRB17_SymmetryEnrichedU1b,
 Benton_PRB20_GroundstatePhaseDiagramb,
 Bhardwaj.Changlani_nQM22_SleuthingOutExotic,
 Desrochers.Kim_PRB23_SymmetryFractionalizationGaugeb,
 Desrochers.Kim_PRL24_SpectroscopicSignaturesFractionalizationb,
 Hosoi.Kim_PRL22_UncoveringFootprintsDipolarOctupolarb}. In particular, a recent neutron scattering and specific heat study \cite{Gao.Dai_NP25_NeutronScatteringThermodynamica} found evidence of an emergent photon in Ce$_2$Zr$_2$O$_7$  consistent with dipolar QSI. Assuming that the monopole gap and bandwidths, $\Delta_\m$ and $W_\m$, and the emergent photon bandwidth, $W_\phi\simeq\hbar c /a$, and magnetic stiffness, $\mu^{-1}$,  are all comparable \footnote{This is reasonable in the strong Ising limit of the QSI, where all these scales are controlled by the ring exchange \cite{Hermele.Balents_PRB04_PyrochlorePhotons$U1$}.}, we estimate its monopole plasma resonance at  $\kB T=0.5\Delta_\m$, would be at $\hbar\omega_p \sim 0.5\hbar c /a \sim 0.0025$meV, where we used the measured photon bandwidth $\hbar c /a \sim 0.005$meV \cite{Gao.Dai_NP25_NeutronScatteringThermodynamica}. Notice that this is an utterly small energy scale compared to the  electron band gap of Ce$_2$Zr$_2$O$_7$ which is about $2.5$eV \cite{Uno.Yamanaka_JAC06_PhotoelectrochemicalStudyLanthanide}.

\begin{figure}[t]
\raisebox{6pt}{\includegraphics[width=0.35\columnwidth,angle=0,clip=true]{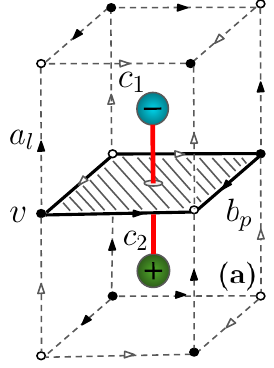}}
\includegraphics[width=0.5\columnwidth,angle=0,clip=true]{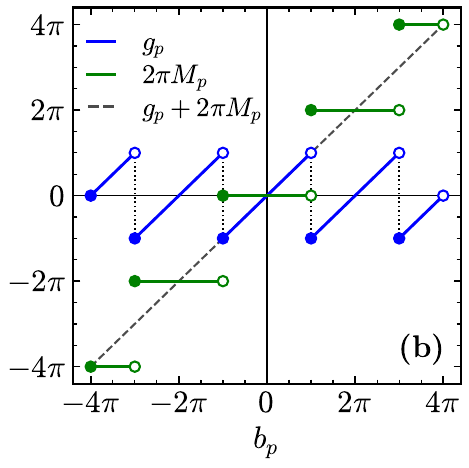}
\caption{ \textbf{(a)} Schematic of monopoles and WGT variables in a cubic lattice.
\textbf{(b)} Decomposition of emergent magnetic field ($b_p$) into "small" part ($g_p$) and Dirac strings ($M_p$).}
\label{fig1}   
\end{figure}

\sectn{Coupling emergent to physical electromagnetic fields}
Our goal is to study a plasma resonance expected in 3D $U(1)$ spin liquids where the emergent magnetic monopole carries physical electric charge, which occurs when the emergent magnetic field, denoted by $b$, has the same transformation under symmetries as the physical electric field $E$ \cite{Laumann.Moessner_PRB23_HybridDyonsInvertedb}.
Based on symmetry analysis, we will argue that this is a  universal phenomenon for dipolar QSI models of 3D $U(1)$ QSLs, but for clarity and simplicity, we begin here by analyzing one the simplest lattice QEDs in 3D: Wilson's gauge theory (WGT) in the cubic lattice \cite{Wilson_PRD74_ConfinementQuarks,Kogut.Susskind_PRD75_HamiltonianFormulationWilsonsb,Kogut_RMP79_IntroductionLatticeGaugea}. 
We will express our key results in terms of effective low energy parameters, so that our formulae can be used to perform estimates in  other microscopic settings \footnote{For microscopic calculations of quasiparticle properties of pyrochlore 3D $U(1)$ QSLs arising in QSI, see e.g. \cite{Hermele.Balents_PRB04_PyrochlorePhotons$U1$,
Gingras.McClarty_RPP14_QuantumSpinIce,
savary_coulombic_2012,
Shannon.Fulde_PRL12_QuantumIceQuantum,
Benton.Shannon_PRB12_SeeingLightExperimental,
Lee.Balents_PRB12_GenericQuantumSpin}
}. WGT is defined in terms of compact gauge fields (rotors), $a_l\in [-\pi,\pi)$ that reside on the links $l$ of a cubic lattice of unit cell length $a$ (see Fig.~\ref{fig1}(a)), through the Lagrangian,
	\begin{align}\label{L0}
	\lagr_0=\frac{1}{2}\epsilon\sum_{l}e_{l}^{2}-\mu^{-1}\sum_{p}1-\cos(b_p),
	\end{align}
where the emergent electric field is $e_l=-\dot{a}_{l}$ and the emergent magnetic flux through a plaquette $p$ is  $b_p=\lcurl_{p}(a_l)$ 
\footnote{The lattice curl is defined as $\lcurl_{p}(a_l)\equiv \sum_{l\in p}\sigma_l^{p}a_{l}$ where $\sigma_l^p$ takes values $\pm1$ along the links that bound the plaquette; see Appendix \ref{appendixA1} for precise definitions of lattice differential operators ($\lgrad,\lcurl,\ldiv$).}. 
Here $\hbar^2/\epsilon$ and $\mu^{-1}$ are the effective stiffness energy scales of the emergent electric and magnetic fields 
\footnote{The theory has a single dimensionless parameter $\kappa=\hbar^{-2}\epsilon/\mu$. For $\kappa \ll 1$ it is in the deconfined Maxwell phase (quantum spin-liquid), and transitions into the confined phase (trivial paramagnet) for $\kappa>\kappa_{\rm crit} \sim 1$.}. 
The theory has a conserved electric charge at each vertex $v$ (analogous to the spin-ice rule), defined as $Q_e(v)=\epsilon \ldiv_v(e_l)$
\footnote{The discrete divergence on a vertex $v$ is $\ldiv_v(e_l)=\sum_{l\in v}\sigma^{v}_{l}e_l$, where $\sigma^{v}_{l}$ takes values $\pm1$. See Appendix \ref{appendixA1} for sign convention.}.
We are interested in low energy phenomena, well below the gap for spin-ice rule violations, and thus assume we are in subspace with $Q_e(v)=0$ for all $v$. The monopole particles can be defined by noting that $b_p$ has a larger compactification radius compared to $a_l$, namely the range $b_p \in [-4\pi,4\pi)$ needs to be viewed as physically distinct configurations. 
This leads to a separation of $b_p=g_{p}+2\pi M_{p}$, into its ``small'' $g_p \in [-\pi,\pi)$ and ``large'' $M_p \in \{-2,-1,0,1,2\}$ components (see Fig.\ref{fig1}(b)). 
Physically, $M_p$ counts the flux of Dirac solenoid strings that pierce plaquette $p$, while $g_{p}$ accounts for the ``smooth'' part of the flux spreading outside the Dirac solenoid strings. 
Since the total flux of $b_p$ through any closed surface vanishes, it follows that:
    \begin{align}
       \ldiv_{c}(g_p)= - \ldiv_c(2\pi M_p) \equiv Q_\m(c),
    \end{align}
where $\ldiv_c(g_p)$ is the dual divergence evaluated at the dual cube center $c$  
\footnote{The discrete divergence at a dual cube center $c$ is $\ldiv_c(g_p)=\sum_{p\in c}\sigma_p^{c}g_{p}$, where $\sigma^{p}_{c}$ takes values $\pm1$. See Appendix \ref{appendixA1} for sign convention.},
and $Q_\m(c)$ is the magnetic monopole charge, which is quantized in multiples of $2\pi$.

To mimic symmetry-allowed coupling of emergent to physical electromagnetic fields in QSI discussed in Ref.\cite{Laumann.Moessner_PRB23_HybridDyonsInvertedb}, we  couple WGT to physical electromagnetic fields $E$ and $B$, via the following extra term in the Lagrangian:
\begin{equation}\label{Ltot}        
	\lagr_{eB}+\lagr_{bE}=
	\geB\sum_{l}e_{l}B_{l}
	+\gbE\sum_{p}\sin(b_{p})E_{p},
\end{equation}
where $\geB$ and $\gbE$ are phenomenological coupling constants.  
Here $E_{p}$ and $B_{l}$ are the physical electric and magnetic fields at center of the plaquette and the mid-point of links respectively. We view them as weak probe  fields for which we are interested in deriving a linear response theory \footnote {Namely they are not viewed as dynamical fields.}.

In the deconfined 3D $U(1)$ QSL phase, the linearized and long wave-length equations of motion become the following Maxwell's equations for the emergent fields coupled to the probe physical electromagnetic fields:
    \begin{align}
        a\,\curl\left[\mu^{-1} g-\gbE\,E\right]&=
        \epsilon\,\dot{e}+\geB\,\dot{B} 
        \label{Amp Maxwell's law}, 
    \\
        a\,\curl{e}&=-\left[\dot{g}+a^2J_\m\right], \label{Faraday's law}
    \\
        a\,\left[\epsilon\,\div{e}+\geB\,\div{B}\right]&=0, \label{Gauss law} 
    \\
        a\,\div{g}&=Q_\m. \label{Magnetic Gauss law}
    \end{align}
where $a$ is the lattice constant, and the monopole current is $J_\m=2\pi a^{-2}\dot{M}_p$. Detailed derivation is provided in Appendix \ref{appendixA1}.
%This is the magnetic monopole current that arises from the propagation/dispersion of the topological excitations $Q_\m$.
%\\
%We can draw a parallel between the \eqref{Gauss law} and $\textbf{D}=\epsilon_0 \textbf{E}+\textbf{P}$ for electric fields in matter.
%By this, we deduce that probe B field (along with $g_{eB}$ as a coupling constant) mimics polarization of the e field.
The speed of emergent light is $c=a/\sqrt{\mu\epsilon}$. Finally, the physical electric current density, $J_\text{phys}$, is the variation of the action $\mathcal{S}=\int\dd{t}\lagr_0+\lagr_{eB}+\lagr_{bE}$ with the external vector potential $A_\text{ext}$, and is given by (see \ref{L_bE} for details):
    \begin{align}
    J_{\text{phys}} \equiv \frac{\delta \mathcal{S}}{\delta A_\text{ext}}=
    \frac{1}{a^3}
    \left(
    \gbE\,\dot{g}
    +
    \geB\curl{e}
    \right). 
    \label{Physical Current}
    \end{align}

%Monopoles are point-like topological defects of the emergent fields and all their properties such as energy gap and dynamical mass, are emergent and can be derived from Eqs.\eqref{L0}-\eqref{Ltot}. As an example,

\noindent \sectn{Forces on Monopoles} We have found that the coupling from Eq.\eqref{Ltot}, implies that in the presence of a smoothly varying  external electric field $E_p$, a monopole of charge $Q_\m$ will experience a force given by:
     \begin{equation}\label{Lorentz force equation}
        F_\m=\frac{Q_\m}{a}
        \left(\frac{g_p}{\mu}-\gbE E_p\right).
     \end{equation} 
where $g_p$ is the ``small'' part of the emergent magnetic field. This equation implies that monopoles carry a physical electric charge $q=-e_\m Q_\m/(2\pi)$, where $e_\m = 2\pi\gbE/a$ will be our unit of electric charge.
This charge is identical to that derived in Ref.\cite{Laumann.Moessner_PRB23_HybridDyonsInvertedb} which used a different approach to derive it. We derived Eq.\eqref{Lorentz force equation} by assuming a slowly varying longitudinal electric $E_p(t)=-\nabla_p(\Phi_c(t))$, where $\Phi_c(t)$ is viewed as a scalar potential evaluated at the dual centers. Then by partial integration the coupling $\lagr^{\Phi}_{bE}$ can be converted into an effective potential energy of the monopole $U^{\Phi}_\m(c)=Q_\m(c)\Phi_c$, which gives rise to the component of the force proportional to $E_p$ in Eq.\eqref{Lorentz force equation} (For details see Appendix \ref{appendixA2}). Similarly, by separating the smooth monopole field $g_p=g_p^o+g_p^m$, into the components created by the monopole of interest, $g_p^o$, and that created by other monopoles $g_p^m$, a similar analysis can be performed for the linearized magnetic energy term $g_p^o\cdot g_p^m/\mu$, which gives rise to the part of the force proportional to $g_p$ in Eq.\eqref{Lorentz force equation}  (See \eqref{Monopole force due to g}).

\sectn{Monopole conductivity} 
At temperatures below their gap $\kB T\ll \Delta_\m$, monopoles of mass $m$ (bandwidth defined as $W_\m=\hbar^2/(ma^2)$) will form a compensated plasma with thermally activated density of positive and negative charges,  $a^3n_\m(T) \thicksim (\kB T/W_\m)^{3/2} \exp (-\Delta_\m/\kB T)$, embedded in a relatively denser gas of emergent photons $a^3n_{\phi}(T) \thicksim (\kB T/W_\phi)^3$
(see Appendix \ref{appendixA3} for details). 
%Here $\Lambda_T$ denotes the thermal wavelength (see Appendix[\ref{appendixA3}] for details). 
In this limit, we can assume that only monopoles with charges $Q_\m=\pm 2\pi$ are thermally activated and behave as test particles immersed in the thermal radiation background of emergent photons \footnote{It is straightforward to generalize our analysis to the case in which several species of monopoles with different charges are thermally activated.}. 
As a result, the monopoles with momentum $\ve{p}$ experience a friction force $-\Gamma\ve{p}$, where $\Gamma$ is their momentum relaxation rate. 
Their current in response to spatially uniform and slowly oscillating external electric field, $E(t)=\Re[E(\omega)\mathrm{e}^{-\ii\omega t}]$, and emergent $g$, from Eq.\eqref{Lorentz force equation}, takes a Drude form \footnote{Within linear response, all physical quantities, e.g. $J_\m$ and $g$, are spatially uniform and oscillate with the same frequency of the external perturbation $E$.},
\begin{equation}\label{Jm}
    J_\m=\sigma_\m(\omega)
    \left(\frac{g}{a\mu}-\frac{\gbE}{a}E\right),
    \quad
    \sigma_\m(\omega)=\frac{n_\m Q_\m^2/m}{\Gamma-\ii\omega}
\end{equation}
where $J_\m=n_\m Q_\m v$ is the  monopole particle number current \footnote{Monopoles of opposite charges will have opposite velocities and their currents add up. Thus $n_\m(T)$ stands here for the net positive density of both positive and negative monopoles.}, and $\sigma_\m(\omega)$ is the monopole conductivity.
%, $D_\m = n_\m Q_\m^2/m$ is the monopole Drude weight, $m$ the monopole mass, and $\omega$ the frequency of the oscillating emergent and physical electric field. 
The above monopole current is {\it not} the physical electric current, $J_\text{phys}$, which is given instead by Eq.\eqref{Physical Current}, and therefore, the monopole conductivity $\sigma_\m(\omega)$ is not the physical electrical conductivity. 
The latter, which we define from $J_\text{phys}=\sigma(\omega) E(\omega)$, can be obtained by re-arranging Eqs.\eqref{Faraday's law},\eqref{Physical Current}, and \eqref{Jm} to get,
    \begin{equation}
        \frac{1}{\sigma(\omega)}
        =
        \frac{1}{(\frac{\gbE}{a})^2\sigma_\m(\omega)}
        +
        \frac{1}{-\ii\omega(\frac{\mu}{a})\left(\frac{\gbE}{a}\right)^2} 
        \label{Ioffe Larkin rule}.
    \end{equation}
The above result is characteristic of a Ioffe-Larkin rule \cite{Ioffe.Larkin_PRB89_GaplessFermionsGaugeb} where the resistivities of a metallic component (monopole plasma) add up with that of an insulating vacuum \footnote{It is interesting to note that we have arrived at this rule by a direct analysis of the dynamics of gauge fields and monopoles without resorting to parton constructions, which are more commonly used to derive it.}.

\sectn{Monopole momentum relaxation rate}
The monopole momentum relaxation rate, $\Gamma$, is a crucial parameter that  controls the width and visibility of the monopole plasma resonance. In the low temperature dilute limit, its temperature dependence is controlled by monopoles  scattering off photons. In this regime where monopoles are slow compared to photons, monopole scattering via emission and absorption of photons is kinematically constrained, and relaxation is dominated by Thompson scattering of monopoles off emergent photons \cite{Oxenius1986}, whose low-energy differential cross-section is given by (see Appendix \ref{appendixA5} for derivation)
\begin{equation}
    \frac{d\sigma_s}{d\Omega}=
    \left(
    \frac{Q_\m^2\mu^{-1}a}{4\pi mc^2}\right)^2
    \frac{1+\cos^2\psi}{2},
\end{equation}
where $\psi$ is the scattering angle. Using a detailed kinetic model of the average momentum transfer per collision from photons to monopoles, we find the low-temperature ($\kB T\ll W_\phi)$ monopole momentum relaxation rate to be (for details see Appendix \ref{appendixA6}):
\begin{equation}
    \Gamma=
    %\frac{2^5\pi^4}{3^35}\frac{Q_\m^4}{m^3c^8}\biggl(\frac{a}{\mu}\biggl)^2\frac{(\kB T)^4}{h^3}. 
    \frac{4\pi\,Q_\m^4}{135}
        \left(\frac{\mu^{-1}}{mc^2}\right)^2
        \frac{\hbar}{ma^2}
        \left(\frac{\kB T}{W_\phi}\right)^4.
    \label{Relaxation rate}
\end{equation}
\begin{figure}[t]
\includegraphics[width=\columnwidth,angle=0]{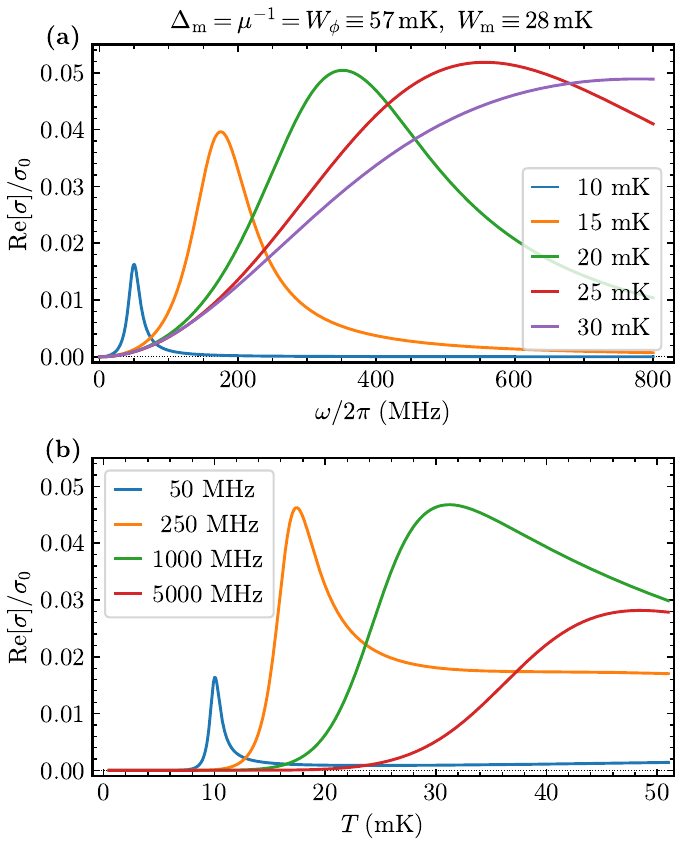}
 \caption{$\Re[\sigma]$ from Eq.~\eqref{sigmaomega} \textbf{(a)} as a function of frequency, and \textbf{(b)} as a function temperature showing a clear monopole plasma resonance peak in both of these traces.
 }
\label{fig:resonance}   
\end{figure}
\sectn{Plasma resonance and dielectric characteristics} Equation \eqref{Ioffe Larkin rule} can be conveniently recast to read,  
% \begin{equation}\label{sigmaomega}
%     \frac{\sigma(\omega)}{\sigma_0}=
%     \frac{\ii \omega_p \omega }{\omega_p^2-\omega^2+\ii \Gamma \omega }, \ \omega_p=\sqrt{\mu^{-1}a D_\m}.
% \end{equation} 
% where $\omega_p=\sqrt{\mu^{-1}a D_\m}$, and $\sigma_0=(g_{bE}/a)^2 D_\m/\omega_p$.  
\begin{equation}\label{sigmaomega}
    \frac{\sigma(\omega)}{\sigma_0}=
    \left(\frac{2\pi\hbar}{ma^2\Gamma}\cdot a^3n_\m\right)
    \frac{\ii \Gamma\omega }%
    {\omega^2-\omega_p^2+\ii \Gamma \omega},
\end{equation} 
where $\sigma_0 = e_\m^2/(ah)$ will be our unit of conductivity, and $\omega_p^2 = n_\m(Q_\m^2\mu^{-1}a)/m$, and using the monopole bandwdith, $W_m=\hbar^2/m a^2$, this  be recast as Eq.\eqref{plasmafreq}. 
%has the dimensions of 3D conductivity of a charge $e_\m$.  
When $\Gamma$ is small ($\Gamma \lesssim \omega_p$), Eq.~\eqref{sigmaomega} displays a sharp plasma-like resonance at a low frequency $\omega\approx\omega_p$ of width $\Gamma$ as illustrated in Fig.\ref{fig:resonance}. This is remarkable because this frequency is expected to be extremely small in comparison to the electron charge gap
% \footnote{This estimates $\omega_p$ from Eq.\eqref{sigmaomega} as the geometric average of an energy scale comparable to the monopole gap, $1/\mu \sim \Delta_\m$, and the energy scale $a D_\m \sim \exp (-\Delta_\m/2k_BT) /m a^2 \sim \Delta_\m \exp (-\Delta_\m/2k_BT) $, where we also assumed that the monopole bandwidth, $1/m a^2$, is comparable to the monopole gap}
, where the $U(1)$ QSL is naively expected to behave as an electrical insulator. 

For quasi-DC response ($\omega\ll\omega_p,\Gamma$) we have $\sigma(\omega)\approx -\ii \omega (\mu g_{bE}^2/a^3)$, which implies that the $U(1)$ spin liquid responds like an insulator under quasi-static external electric fields, despite containing thermally excited itinerant and electrically charged monopoles. In Appendix \ref{appendixA4} we provide an alternative way to understand this quasi-static behavior by computing the polarization of a slab of the $U(1)$ QSL placed in an external uniform electric field $E_0$, and showing that the field inside the slab, $E_\text{in}$ is reduced by monopoles screening at the boundary of the slab as follows $E_\text{in}=E_0/(1+\chi_{\epsilon}), \ \chi_{\epsilon}=(\gbE/a)^2/(\epsilon_0\mu^{-1}a)$. Using the standard relation between dielectric susceptibility $\chi_{\epsilon}$ and conductivity of insulators (see Appendix \ref{appendixA4}), $\sigma(\omega)=-\ii\omega\epsilon_0\chi_{\epsilon}$, we self-consistently recover the low frequency form of the conductivity from Eq.\eqref{sigmaomega}.

When the relaxation rate $\Gamma$ is large ($\Gamma\gg\omega_p$), there might not be resonant peak as a function of frequency. But we find that there is a non-trivial rapid onset of the conductivity around $\omega \sim \omega_p^2/\Gamma$, and a non-monotonic dependence of the conductivity as a function of temperature, that can be used to pinpoint the presence of the monopole plasma in experiments (see Appendix \ref{ap:regimes} for details).

\sectn{Monopole charge in Dipolar vs. Octupolar QSI}
As argued in the seminal work of Ref.\cite{Huang.Hermele_PRL14_QuantumSpinIcesb}, certain compounds (such as Ce$_2$Zr$_2$O$_7$) feature pseudo-spin 1/2 degrees of freedom at pyrochlore sites, that depending on the dominant  anisotropy, have easy-axis components that transform either as ordinary magnetic dipoles (D), or as magnetic octupoles (O). We now address, whether the emergent monopoles from such DQSI or OQSI spin liquids would be electrically charged, and thus, whether the monopole plasma resonance  is expected in these cases. A symmetry analysis (detailed in Appendix \ref{appendixA7}), leads to the transformation rules in Table \ref{table1} for the emergent electric ($\eme_D$,$\eme_O$) and magnetic fields ($\emg_D$,$\emg_O$) in dipolar and octupolar cases. The coupling introduced in Eq.\eqref{Ltot} can only happen if the emergent magnetic field ($\mathbf{g}$) has the same transformation laws as the physical electric field (${\bf E}$). Interestingly, we see from Table \ref{table1} that the mirror symmetry $\Msym$ (see Ref.\cite{Huang.Hermele_PRL14_QuantumSpinIcesb}), forbids this coupling in the octupolar (O) case, and allows it in dipolar (D) case. Therefore, if this symmetry is present, we expect the monopoles of DQSI to be electrically charged, while those of OQSI to be charge neutral \footnote{Notice that emergent electric associated with spin-ice violation charges are always neutral, e.g. by time-reversal or inversion symmetries.}, and thus, the monopole plasma resonance would be only expected in the DQSI and not in OQSI. Therefore, the monopole plasma resonance can distinguish these two kinds of spin liquids experimentally.

\begin{table}[t]
\caption{Transformation rules for emergent and physical fields in dipolar (D) and octupolar (O) pyrochlore QSI. }
\label{table1}
\begin{center}
\renewcommand{\arraystretch}{1.18}
\begin{tabular}{lccc}
\toprule
Field & Inversion \(\Id\) & Mirror \(\Msym\) & Time reversal \(\cT\) \\
\midrule
\(\eme_D\) & pseudo-vector & pseudo-vector & odd \\
\(\emg_D\) & vector & vector & even \\
\(\eme_O\) & pseudo-vector & vector & odd \\
\(\emg_O\) & vector & pseudo-vector & even \\
\(\physE\) & vector & vector & even \\
\(\physB\) & pseudo-vector & pseudo-vector & odd \\
\bottomrule
\end{tabular}
\end{center}
\end{table}

\sectn{Discussion}
Let us discuss the experimental signatures and optimal conditions to observe the monopole plasma resonance in QSI materials. 
A distinctive characteristic is that the resonant frequency shifts with temperature (see Fig.~\ref{fig:resonance} and Eq.~\eqref{plasmafreq}): the lower the temperature, the lower the density of thermally activated monopoles, leading to a lower resonant frequency. The height of the resonant peak also changes with temperature (see Fig.~\ref{fig:resonance}(a)) in a non-monotonic fashion, and peaks at roughly $40\%$ of the monopole gap. 
Therefore, experiments should reach temperatures below the monopole gap 
\footnote{This is already achieved for Ce$_2$Zr$_2$O$_7$ in e.g. Ref.\cite{Gao.Dai_NP25_NeutronScatteringThermodynamica}, since at their lowest temperatures the specific heat is dominated by emergent photons. In optical experiments it will be important to monitor that there is no light induced heating that could substantially raise the temperature of spins.}. 
Also, for octupolar QSI the monopoles are charge neutral, and thus the QSI should be dipolar, which is also believed for Ce$_2$Zr$_2$O$_7$ based on its neutron scattering characteristics \cite{Gao.Dai_NP25_NeutronScatteringThermodynamica}. 
Therefore, Ce$_2$Zr$_2$O$_7$ is a promising material for observing the monopole plasma resonance. 

Another useful feature is that the resonance can also be detected by sweeping temperatures at fixed frequency (see Fig.\ref{fig:resonance}(b)). 
If we assume that the monopole gap and magnetic stiffness scales are comparable to the measured photon bandwidth and the monopole bandwidth is $W_\m=0.5W_\phi$, driving the system at a frequency $2\pi\omega=1000\,\text{MHz}$ (comparable to $W_\phi/\hbar\sim1.2\,\text{GHz}$ \cite{Gao.Dai_NP25_NeutronScatteringThermodynamica}) results in a peak in conductivity at $T^\ast\approx31\,\text{mK}$ (green trace in Fig.\ref{fig:resonance}(b)). 
Driving at a lower frequency, $2\pi\omega=250\,\text{MHz}$, shifts the peak in conductivity to a lower temperature $T^\ast\approx 17\,\text{mK}$ and makes the peak {\it sharper} (orange trace in Fig.\ref{fig:resonance}(b))
\footnote{If instead we assume that {\it all} the monopole energy scales are comparable to the photon bandwidth, 
the conductivity changes with temperature in an almost similar fashion at different driving frequencies (see Appendix \ref{ap:regimes}). 
The conductivity peaks at $T^\ast\approx 25\,\text{mK}$, and the maximum value at the peak becomes $\sigma_\text{max}\approx0.005\,e_\m^2/(ah)$, which is somewhat smaller, but still sizable.}.

% we expect its monopole resonance to be around $f_p=2 \pi \omega_p\simeq 0.6$GHz at a temperature of $T\approx 25$mK. Given the substantial uncertainties on microscopic details of monopole properties, it would be desirable to experimentally cover at least one order of magnitude in frequency ranges around this value to increase the chances of observing this resonance \footnote{For example, if we keep the temperature at half the monopole gap, but instead take the monopole band-width to be either 4 times larger and 4 times smaller than the monopole gap, the resonance would shift $1.7$ GHz and $0.2$ GHz respectively.}. 
The maximum conductivity at such a peak is $\sigma_\text{max}\approx0.05\,e_\m^2/(ah)$, where $e_\m$ is the monopole electric charge, which is  a parameter that is challenging to estimate microscopically, but which is expected to be smaller than the electron charge. 
A useful comparison is to consider the measured DC resistivity of stoichiometric Ce$_2$Zr$_2$O$_7$ (lattice constant $a= 10.7\,\AA$ \cite{Gao.Dai_NP19_ExperimentalSignaturesThreedimensionalb}) at room temperature (RT), $\rho \sim 10^7 \Omega\,\text{cm}$  \cite{Baidya.Gopalakrishnan_JPCB07_OxygenReleaseStoragePropertiesa}, which leads to a conductivity, $\sigma_\text{RT} \sim 2.8 \times 10^{-10} e^2/(ah)$.  
So, the ratio of the monopole plasma peak conductivity to the RT conductivity is $\sigma_\text{max}/\sigma_\text{RT} \sim 1.8 \times 10^{8} \times (e_\m/e)^2$. 
In other words, even if the charge of the monopoles in Ce$_2$Zr$_2$O$_7$ is $10^{-4}$ times smaller than the electron's, the peak conductivity would be higher than its DC conductivity at room temperature, making us optimistic that the monopole resonance could be visible in experiments. 

We close by re-iterating that the monopoles being charged in these QSLs is a highly non-trivial physical property. We have focused here on its implications for the frequency dependent electrical conductivity, but this amusing property should have important consequences in many other properties and observables \cite{Laumann.Moessner_PRB23_HybridDyonsInvertedb}. One interesting direction for future investigations is to understand its interplay with lattice deformations and phonons.

% This is a high density of charges, even if $g_{bE}$ is small, it can overwhelm the E\&M response, analogous to how these degrees of freedom overwhelm the specific heat. This sharp distinction between monopoles havin monopole has charge, there should be strong monopole coupling to elastic deformations of unit cells (phonons). A topic for future investigations.

\sectn{Acknowledgements} 
We are thankful to Silke Bühler-Paschen, Roderich Moessner, Chris Laumann, Jeffrey Rau, Yong-Baek Kim, and Michael Hermele for valuable discussions. We acknowledge support by the Deutsche Forschungsgemeinschaft (DFG) through research grant project numbers 542614019; 518372354; 555335098.

\bibliographystyle{apsrev4-2}
\bibliography{monopoleplasma}

\begin{thebibliography}{65}%
\makeatletter
\providecommand \@ifxundefined [1]{%
 \@ifx{#1\undefined}
}%
\providecommand \@ifnum [1]{%
 \ifnum #1\expandafter \@firstoftwo
 \else \expandafter \@secondoftwo
 \fi
}%
\providecommand \@ifx [1]{%
 \ifx #1\expandafter \@firstoftwo
 \else \expandafter \@secondoftwo
 \fi
}%
\providecommand \natexlab [1]{#1}%
\providecommand \enquote  [1]{``#1''}%
\providecommand \bibnamefont  [1]{#1}%
\providecommand \bibfnamefont [1]{#1}%
\providecommand \citenamefont [1]{#1}%
\providecommand \href@noop [0]{\@secondoftwo}%
\providecommand \href [0]{\begingroup \@sanitize@url \@href}%
\providecommand \@href[1]{\@@startlink{#1}\@@href}%
\providecommand \@@href[1]{\endgroup#1\@@endlink}%
\providecommand \@sanitize@url [0]{\catcode `\\12\catcode `\$12\catcode
  `\&12\catcode `\#12\catcode `\^12\catcode `\_12\catcode `\%12\relax}%
\providecommand \@@startlink[1]{}%
\providecommand \@@endlink[0]{}%
\providecommand \url  [0]{\begingroup\@sanitize@url \@url }%
\providecommand \@url [1]{\endgroup\@href {#1}{\urlprefix }}%
\providecommand \urlprefix  [0]{URL }%
\providecommand \Eprint [0]{\href }%
\providecommand \doibase [0]{https://doi.org/}%
\providecommand \selectlanguage [0]{\@gobble}%
\providecommand \bibinfo  [0]{\@secondoftwo}%
\providecommand \bibfield  [0]{\@secondoftwo}%
\providecommand \translation [1]{[#1]}%
\providecommand \BibitemOpen [0]{}%
\providecommand \bibitemStop [0]{}%
\providecommand \bibitemNoStop [0]{.\EOS\space}%
\providecommand \EOS [0]{\spacefactor3000\relax}%
\providecommand \BibitemShut  [1]{\csname bibitem#1\endcsname}%
\let\auto@bib@innerbib\@empty
%</preamble>
\bibitem [{\citenamefont {Wen}(2001)}]{Wen_PRL01_OriginGaugeBosons}%
  \BibitemOpen
  \bibfield  {author} {\bibinfo {author} {\bibfnamefont {X.-G.}\ \bibnamefont
  {Wen}},\ }\href {https://doi.org/10.1103/PhysRevLett.88.011602} {\bibfield
  {journal} {\bibinfo  {journal} {Physical Review Letters}\ }\textbf {\bibinfo
  {volume} {88}},\ \bibinfo {pages} {011602} (\bibinfo {year}
  {2001})}\BibitemShut {NoStop}%
\bibitem [{\citenamefont {Hermele}\ \emph {et~al.}(2004)\citenamefont
  {Hermele}, \citenamefont {Fisher},\ and\ \citenamefont
  {Balents}}]{Hermele.Balents_PRB04_PyrochlorePhotons$U1$}%
  \BibitemOpen
  \bibfield  {author} {\bibinfo {author} {\bibfnamefont {M.}~\bibnamefont
  {Hermele}}, \bibinfo {author} {\bibfnamefont {M.~P.~A.}\ \bibnamefont
  {Fisher}},\ and\ \bibinfo {author} {\bibfnamefont {L.}~\bibnamefont
  {Balents}},\ }\href {https://doi.org/10.1103/PhysRevB.69.064404} {\bibfield
  {journal} {\bibinfo  {journal} {Physical Review B}\ }\textbf {\bibinfo
  {volume} {69}},\ \bibinfo {pages} {064404} (\bibinfo {year}
  {2004})}\BibitemShut {NoStop}%
\bibitem [{\citenamefont {Gingras}\ and\ \citenamefont
  {McClarty}(2014)}]{Gingras.McClarty_RPP14_QuantumSpinIce}%
  \BibitemOpen
  \bibfield  {author} {\bibinfo {author} {\bibfnamefont {M.~J.~P.}\
  \bibnamefont {Gingras}}\ and\ \bibinfo {author} {\bibfnamefont {P.~A.}\
  \bibnamefont {McClarty}},\ }\href
  {https://doi.org/10.1088/0034-4885/77/5/056501} {\bibfield  {journal}
  {\bibinfo  {journal} {Reports on Progress in Physics}\ }\textbf {\bibinfo
  {volume} {77}},\ \bibinfo {pages} {056501} (\bibinfo {year} {2014})},\
  \bibinfo {note} {arXiv:1311.1817 [cond-mat.str-el]}\BibitemShut {NoStop}%
\bibitem [{\citenamefont {Motrunich}\ and\ \citenamefont
  {Senthil}(2002)}]{Motrunich.Senthil_PRL02_ExoticOrderSimple}%
  \BibitemOpen
  \bibfield  {author} {\bibinfo {author} {\bibfnamefont {O.~I.}\ \bibnamefont
  {Motrunich}}\ and\ \bibinfo {author} {\bibfnamefont {T.}~\bibnamefont
  {Senthil}},\ }\href {https://doi.org/10.1103/PhysRevLett.89.277004}
  {\bibfield  {journal} {\bibinfo  {journal} {Physical Review Letters}\
  }\textbf {\bibinfo {volume} {89}},\ \bibinfo {pages} {277004} (\bibinfo
  {year} {2002})}\BibitemShut {NoStop}%
\bibitem [{\citenamefont {Senthil}\ and\ \citenamefont
  {Motrunich}(2002)}]{Senthil.Motrunich_PRB02_MicroscopicModelsFractionalized}%
  \BibitemOpen
  \bibfield  {author} {\bibinfo {author} {\bibfnamefont {T.}~\bibnamefont
  {Senthil}}\ and\ \bibinfo {author} {\bibfnamefont {O.}~\bibnamefont
  {Motrunich}},\ }\href {https://doi.org/10.1103/PhysRevB.66.205104} {\bibfield
   {journal} {\bibinfo  {journal} {Physical Review B}\ }\textbf {\bibinfo
  {volume} {66}},\ \bibinfo {pages} {205104} (\bibinfo {year}
  {2002})}\BibitemShut {NoStop}%
\bibitem [{\citenamefont {Wen}(2003)}]{Wen_PRB03_ArtificialLightQuantum}%
  \BibitemOpen
  \bibfield  {author} {\bibinfo {author} {\bibfnamefont {X.-G.}\ \bibnamefont
  {Wen}},\ }\href {https://doi.org/10.1103/PhysRevB.68.115413} {\bibfield
  {journal} {\bibinfo  {journal} {Physical Review B}\ }\textbf {\bibinfo
  {volume} {68}},\ \bibinfo {pages} {115413} (\bibinfo {year}
  {2003})}\BibitemShut {NoStop}%
\bibitem [{\citenamefont {Banerjee}\ \emph {et~al.}(2008)\citenamefont
  {Banerjee}, \citenamefont {Isakov}, \citenamefont {Damle},\ and\
  \citenamefont {Kim}}]{banerjee_unusual_2008}%
  \BibitemOpen
  \bibfield  {author} {\bibinfo {author} {\bibfnamefont {A.}~\bibnamefont
  {Banerjee}}, \bibinfo {author} {\bibfnamefont {S.~V.}\ \bibnamefont
  {Isakov}}, \bibinfo {author} {\bibfnamefont {K.}~\bibnamefont {Damle}},\ and\
  \bibinfo {author} {\bibfnamefont {Y.~B.}\ \bibnamefont {Kim}},\ }\href
  {https://doi.org/10.1103/PhysRevLett.100.047208} {\bibfield  {journal}
  {\bibinfo  {journal} {Physical Review Letters}\ }\textbf {\bibinfo {volume}
  {100}},\ \bibinfo {pages} {047208} (\bibinfo {year} {2008})}\BibitemShut
  {NoStop}%
\bibitem [{Note1()}]{Note1}%
  \BibitemOpen
  \bibinfo {note} {Provided that it is local (e.g. only couples a finite number
  of neighboring spins) and it is not too strong so as to induce a phase
  transition into another state of matter.}\BibitemShut {Stop}%
\bibitem [{Note2()}]{Note2}%
  \BibitemOpen
  \bibinfo {note} {We are using the terminology of Ref.\cite
  {Hermele.Balents_PRB04_PyrochlorePhotons$U1$}, where the spin-ice rule
  violations would be viewed the emergent electric charges (i.e. not the
  magnetic monopoles). The magnetic monopoles have also been referred to as
  visons \cite {Gingras.McClarty_RPP14_QuantumSpinIce}.}\BibitemShut {Stop}%
\bibitem [{Note3()}]{Note3}%
  \BibitemOpen
  \bibinfo {note} {Namely their energy cost approaches a constant in the limit
  of long strings}\BibitemShut {NoStop}%
\bibitem [{\citenamefont
  {Villadiego}(2021)}]{Villadiego_PRB21_PseudoscalarU1Spin}%
  \BibitemOpen
  \bibfield  {author} {\bibinfo {author} {\bibfnamefont {I.~S.}\ \bibnamefont
  {Villadiego}},\ }\href {https://doi.org/10.1103/PhysRevB.104.195149}
  {\bibfield  {journal} {\bibinfo  {journal} {Physical Review B}\ }\textbf
  {\bibinfo {volume} {104}},\ \bibinfo {pages} {195149} (\bibinfo {year}
  {2021})}\BibitemShut {NoStop}%
\bibitem [{\citenamefont {Ioffe}\ and\ \citenamefont
  {Larkin}(1989)}]{Ioffe.Larkin_PRB89_GaplessFermionsGaugeb}%
  \BibitemOpen
  \bibfield  {author} {\bibinfo {author} {\bibfnamefont {L.~B.}\ \bibnamefont
  {Ioffe}}\ and\ \bibinfo {author} {\bibfnamefont {A.~I.}\ \bibnamefont
  {Larkin}},\ }\href {https://doi.org/10.1103/PhysRevB.39.8988} {\bibfield
  {journal} {\bibinfo  {journal} {Physical Review B}\ }\textbf {\bibinfo
  {volume} {39}},\ \bibinfo {pages} {8988} (\bibinfo {year}
  {1989})}\BibitemShut {NoStop}%
\bibitem [{\citenamefont
  {Motrunich}(2006)}]{Motrunich_PRB06_OrbitalMagneticField}%
  \BibitemOpen
  \bibfield  {author} {\bibinfo {author} {\bibfnamefont {O.~I.}\ \bibnamefont
  {Motrunich}},\ }\href {https://doi.org/10.1103/PhysRevB.73.155115} {\bibfield
   {journal} {\bibinfo  {journal} {Physical Review B}\ }\textbf {\bibinfo
  {volume} {73}},\ \bibinfo {pages} {155115} (\bibinfo {year}
  {2006})}\BibitemShut {NoStop}%
\bibitem [{\citenamefont {Ng}\ and\ \citenamefont
  {Lee}(2007)}]{Ng.Lee_PRL07_PowerLawConductivityMott}%
  \BibitemOpen
  \bibfield  {author} {\bibinfo {author} {\bibfnamefont {T.-K.}\ \bibnamefont
  {Ng}}\ and\ \bibinfo {author} {\bibfnamefont {P.~A.}\ \bibnamefont {Lee}},\
  }\href {https://doi.org/10.1103/PhysRevLett.99.156402} {\bibfield  {journal}
  {\bibinfo  {journal} {Physical Review Letters}\ }\textbf {\bibinfo {volume}
  {99}},\ \bibinfo {pages} {156402} (\bibinfo {year} {2007})}\BibitemShut
  {NoStop}%
\bibitem [{\citenamefont {Bulaevskii}\ \emph {et~al.}(2008)\citenamefont
  {Bulaevskii}, \citenamefont {Batista}, \citenamefont {Mostovoy},\ and\
  \citenamefont
  {Khomskii}}]{Bulaevskii.Khomskii_PRB08_ElectronicOrbitalCurrents}%
  \BibitemOpen
  \bibfield  {author} {\bibinfo {author} {\bibfnamefont {L.~N.}\ \bibnamefont
  {Bulaevskii}}, \bibinfo {author} {\bibfnamefont {C.~D.}\ \bibnamefont
  {Batista}}, \bibinfo {author} {\bibfnamefont {M.}~\bibnamefont {Mostovoy}},\
  and\ \bibinfo {author} {\bibfnamefont {D.}~\bibnamefont {Khomskii}},\ }\href
  {https://doi.org/10.1103/PhysRevB.78.024402} {\bibfield  {journal} {\bibinfo
  {journal} {Physical Review B}\ }\textbf {\bibinfo {volume} {78}},\ \bibinfo
  {pages} {024402} (\bibinfo {year} {2008})},\ \bibinfo {note} {arXiv:0709.0575
  [cond-mat.str-el]}\BibitemShut {NoStop}%
\bibitem [{\citenamefont
  {Khomskii}(2012)}]{Khomskii_NC12_ElectricDipolesMagnetic}%
  \BibitemOpen
  \bibfield  {author} {\bibinfo {author} {\bibfnamefont {D.~I.}\ \bibnamefont
  {Khomskii}},\ }\href {https://doi.org/10.1038/ncomms1904} {\bibfield
  {journal} {\bibinfo  {journal} {Nature Communications}\ }\textbf {\bibinfo
  {volume} {3}},\ \bibinfo {pages} {904} (\bibinfo {year} {2012})},\ \bibinfo
  {note} {arXiv:1206.0402 [cond-mat.str-el]}\BibitemShut {NoStop}%
\bibitem [{\citenamefont {Potter}\ \emph {et~al.}(2013)\citenamefont {Potter},
  \citenamefont {Senthil},\ and\ \citenamefont
  {Lee}}]{Potter.Lee_PRB13_MechanismsSubgapOptical}%
  \BibitemOpen
  \bibfield  {author} {\bibinfo {author} {\bibfnamefont {A.~C.}\ \bibnamefont
  {Potter}}, \bibinfo {author} {\bibfnamefont {T.}~\bibnamefont {Senthil}},\
  and\ \bibinfo {author} {\bibfnamefont {P.~A.}\ \bibnamefont {Lee}},\ }\href
  {https://doi.org/10.1103/PhysRevB.87.245106} {\bibfield  {journal} {\bibinfo
  {journal} {Physical Review B}\ }\textbf {\bibinfo {volume} {87}},\ \bibinfo
  {pages} {245106} (\bibinfo {year} {2013})}\BibitemShut {NoStop}%
\bibitem [{\citenamefont {Lantagne-Hurtubise}\ \emph
  {et~al.}(2017)\citenamefont {Lantagne-Hurtubise}, \citenamefont
  {Bhattacharjee},\ and\ \citenamefont
  {Moessner}}]{Lantagne-Hurtubise.Moessner_PRB17_ElectricFieldControl}%
  \BibitemOpen
  \bibfield  {author} {\bibinfo {author} {\bibfnamefont {E.}~\bibnamefont
  {Lantagne-Hurtubise}}, \bibinfo {author} {\bibfnamefont {S.}~\bibnamefont
  {Bhattacharjee}},\ and\ \bibinfo {author} {\bibfnamefont {R.}~\bibnamefont
  {Moessner}},\ }\href {https://doi.org/10.1103/PhysRevB.96.125145} {\bibfield
  {journal} {\bibinfo  {journal} {Physical Review B}\ }\textbf {\bibinfo
  {volume} {96}},\ \bibinfo {pages} {125145} (\bibinfo {year}
  {2017})}\BibitemShut {NoStop}%
\bibitem [{\citenamefont {Fu}\ \emph {et~al.}(2017)\citenamefont {Fu},
  \citenamefont {Rau}, \citenamefont {Gingras},\ and\ \citenamefont
  {Perkins}}]{Fu.Perkins_PRB17_FingerprintsQuantumSpin}%
  \BibitemOpen
  \bibfield  {author} {\bibinfo {author} {\bibfnamefont {J.}~\bibnamefont
  {Fu}}, \bibinfo {author} {\bibfnamefont {J.~G.}\ \bibnamefont {Rau}},
  \bibinfo {author} {\bibfnamefont {M.~J.~P.}\ \bibnamefont {Gingras}},\ and\
  \bibinfo {author} {\bibfnamefont {N.~B.}\ \bibnamefont {Perkins}},\ }\href
  {https://doi.org/10.1103/PhysRevB.96.035136} {\bibfield  {journal} {\bibinfo
  {journal} {Physical Review B}\ }\textbf {\bibinfo {volume} {96}},\ \bibinfo
  {pages} {035136} (\bibinfo {year} {2017})},\ \bibinfo {note}
  {arXiv:1703.03836 [cond-mat.str-el]}\BibitemShut {NoStop}%
\bibitem [{\citenamefont {Sodemann}\ \emph {et~al.}(2018)\citenamefont
  {Sodemann}, \citenamefont {Chowdhury},\ and\ \citenamefont
  {Senthil}}]{Sodemann.Senthil_PRB18_QuantumOscillationsInsulators}%
  \BibitemOpen
  \bibfield  {author} {\bibinfo {author} {\bibfnamefont {I.}~\bibnamefont
  {Sodemann}}, \bibinfo {author} {\bibfnamefont {D.}~\bibnamefont
  {Chowdhury}},\ and\ \bibinfo {author} {\bibfnamefont {T.}~\bibnamefont
  {Senthil}},\ }\href {https://doi.org/10.1103/PhysRevB.97.045152} {\bibfield
  {journal} {\bibinfo  {journal} {Physical Review B}\ }\textbf {\bibinfo
  {volume} {97}},\ \bibinfo {pages} {045152} (\bibinfo {year}
  {2018})}\BibitemShut {NoStop}%
\bibitem [{\citenamefont {Rao}\ and\ \citenamefont
  {Sodemann}(2019)}]{Rao.Sodemann_PRB19_CyclotronResonanceMott}%
  \BibitemOpen
  \bibfield  {author} {\bibinfo {author} {\bibfnamefont {P.}~\bibnamefont
  {Rao}}\ and\ \bibinfo {author} {\bibfnamefont {I.}~\bibnamefont {Sodemann}},\
  }\href {https://doi.org/10.1103/PhysRevB.100.155150} {\bibfield  {journal}
  {\bibinfo  {journal} {Physical Review B}\ }\textbf {\bibinfo {volume}
  {100}},\ \bibinfo {pages} {155150} (\bibinfo {year} {2019})}\BibitemShut
  {NoStop}%
\bibitem [{\citenamefont {Khoo}\ \emph
  {et~al.}(2022{\natexlab{a}})\citenamefont {Khoo}, \citenamefont {Pientka},\
  and\ \citenamefont
  {Sodemann~Villadiego}}]{Khoo.SodemannVilladiego_NJP22_CorrigendumUniversalShear}%
  \BibitemOpen
  \bibfield  {author} {\bibinfo {author} {\bibfnamefont {J.~Y.}\ \bibnamefont
  {Khoo}}, \bibinfo {author} {\bibfnamefont {F.}~\bibnamefont {Pientka}},\ and\
  \bibinfo {author} {\bibfnamefont {I.}~\bibnamefont {Sodemann~Villadiego}},\
  }\href {https://doi.org/10.1088/1367-2630/acac46} {\bibfield  {journal}
  {\bibinfo  {journal} {New Journal of Physics}\ }\textbf {\bibinfo {volume}
  {24}},\ \bibinfo {pages} {129501} (\bibinfo {year}
  {2022}{\natexlab{a}})}\BibitemShut {NoStop}%
\bibitem [{\citenamefont {Khoo}\ \emph
  {et~al.}(2022{\natexlab{b}})\citenamefont {Khoo}, \citenamefont {Pientka},
  \citenamefont {Lee},\ and\ \citenamefont
  {Villadiego}}]{Khoo.Villadiego_PRB22_ProbingQuantumNoise}%
  \BibitemOpen
  \bibfield  {author} {\bibinfo {author} {\bibfnamefont {J.~Y.}\ \bibnamefont
  {Khoo}}, \bibinfo {author} {\bibfnamefont {F.}~\bibnamefont {Pientka}},
  \bibinfo {author} {\bibfnamefont {P.~A.}\ \bibnamefont {Lee}},\ and\ \bibinfo
  {author} {\bibfnamefont {I.~S.}\ \bibnamefont {Villadiego}},\ }\href
  {https://doi.org/10.1103/PhysRevB.106.115108} {\bibfield  {journal} {\bibinfo
   {journal} {Physical Review B}\ }\textbf {\bibinfo {volume} {106}},\ \bibinfo
  {pages} {115108} (\bibinfo {year} {2022}{\natexlab{b}})}\BibitemShut
  {NoStop}%
\bibitem [{\citenamefont {Laumann}\ and\ \citenamefont
  {Moessner}(2023)}]{Laumann.Moessner_PRB23_HybridDyonsInvertedb}%
  \BibitemOpen
  \bibfield  {author} {\bibinfo {author} {\bibfnamefont {C.~R.}\ \bibnamefont
  {Laumann}}\ and\ \bibinfo {author} {\bibfnamefont {R.}~\bibnamefont
  {Moessner}},\ }\href {https://doi.org/10.1103/PhysRevB.108.L220402}
  {\bibfield  {journal} {\bibinfo  {journal} {Physical Review B}\ }\textbf
  {\bibinfo {volume} {108}},\ \bibinfo {pages} {L220402} (\bibinfo {year}
  {2023})}\BibitemShut {NoStop}%
\bibitem [{\citenamefont {Wu}\ \emph {et~al.}(2024)\citenamefont {Wu},
  \citenamefont {Schoop}, \citenamefont {Sodemann}, \citenamefont {Moessner},
  \citenamefont {Cava},\ and\ \citenamefont
  {Ong}}]{Wu.Ong_N24_ChargeNeutralElectronicExcitations}%
  \BibitemOpen
  \bibfield  {author} {\bibinfo {author} {\bibfnamefont {S.}~\bibnamefont
  {Wu}}, \bibinfo {author} {\bibfnamefont {L.~M.}\ \bibnamefont {Schoop}},
  \bibinfo {author} {\bibfnamefont {I.}~\bibnamefont {Sodemann}}, \bibinfo
  {author} {\bibfnamefont {R.}~\bibnamefont {Moessner}}, \bibinfo {author}
  {\bibfnamefont {R.~J.}\ \bibnamefont {Cava}},\ and\ \bibinfo {author}
  {\bibfnamefont {N.~P.}\ \bibnamefont {Ong}},\ }\href
  {https://doi.org/10.1038/s41586-024-08091-8} {\bibfield  {journal} {\bibinfo
  {journal} {Nature}\ }\textbf {\bibinfo {volume} {635}},\ \bibinfo {pages}
  {301} (\bibinfo {year} {2024})},\ \bibinfo {note} {arXiv:2411.09496
  [cond-mat.str-el]}\BibitemShut {NoStop}%
\bibitem [{Note4()}]{Note4}%
  \BibitemOpen
  \bibinfo {note} {Namely the energy scale obtained from the coefficient in
  front of the Maxwell-like term associated with the low energy cost for
  emergent magnetic fields, which in QSI is controlled by the ring exchange
  energy scale \cite {Hermele.Balents_PRB04_PyrochlorePhotons$U1$} and is also
  typically comparable to the monopole gap $\Delta _\protect \text
  {m}$.}\BibitemShut {Stop}%
\bibitem [{\citenamefont {Gaudet}\ \emph {et~al.}(2019)\citenamefont {Gaudet},
  \citenamefont {Smith}, \citenamefont {Dudemaine}, \citenamefont {Beare},
  \citenamefont {Buhariwalla}, \citenamefont {Butch}, \citenamefont {Stone},
  \citenamefont {Kolesnikov}, \citenamefont {Xu}, \citenamefont {Yahne},
  \citenamefont {Ross}, \citenamefont {Marjerrison}, \citenamefont {Garrett},
  \citenamefont {Luke}, \citenamefont {Bianchi},\ and\ \citenamefont
  {Gaulin}}]{Gaudet.Gaulin_PRL19_QuantumSpinIceb}%
  \BibitemOpen
  \bibfield  {author} {\bibinfo {author} {\bibfnamefont {J.}~\bibnamefont
  {Gaudet}}, \bibinfo {author} {\bibfnamefont {E.}~\bibnamefont {Smith}},
  \bibinfo {author} {\bibfnamefont {J.}~\bibnamefont {Dudemaine}}, \bibinfo
  {author} {\bibfnamefont {J.}~\bibnamefont {Beare}}, \bibinfo {author}
  {\bibfnamefont {C.}~\bibnamefont {Buhariwalla}}, \bibinfo {author}
  {\bibfnamefont {N.}~\bibnamefont {Butch}}, \bibinfo {author} {\bibfnamefont
  {M.}~\bibnamefont {Stone}}, \bibinfo {author} {\bibfnamefont
  {A.}~\bibnamefont {Kolesnikov}}, \bibinfo {author} {\bibfnamefont
  {G.}~\bibnamefont {Xu}}, \bibinfo {author} {\bibfnamefont {D.}~\bibnamefont
  {Yahne}}, \bibinfo {author} {\bibfnamefont {K.}~\bibnamefont {Ross}},
  \bibinfo {author} {\bibfnamefont {C.}~\bibnamefont {Marjerrison}}, \bibinfo
  {author} {\bibfnamefont {J.}~\bibnamefont {Garrett}}, \bibinfo {author}
  {\bibfnamefont {G.}~\bibnamefont {Luke}}, \bibinfo {author} {\bibfnamefont
  {A.}~\bibnamefont {Bianchi}},\ and\ \bibinfo {author} {\bibfnamefont
  {B.}~\bibnamefont {Gaulin}},\ }\href
  {https://doi.org/10.1103/PhysRevLett.122.187201} {\bibfield  {journal}
  {\bibinfo  {journal} {Physical Review Letters}\ }\textbf {\bibinfo {volume}
  {122}},\ \bibinfo {pages} {187201} (\bibinfo {year} {2019})}\BibitemShut
  {NoStop}%
\bibitem [{\citenamefont {Gao}\ \emph {et~al.}(2019)\citenamefont {Gao},
  \citenamefont {Chen}, \citenamefont {Tam}, \citenamefont {Huang},
  \citenamefont {Sasmal}, \citenamefont {Adroja}, \citenamefont {Ye},
  \citenamefont {Cao}, \citenamefont {Sala}, \citenamefont {Stone},
  \citenamefont {Baines}, \citenamefont {Verezhak}, \citenamefont {Hu},
  \citenamefont {Chung}, \citenamefont {Xu}, \citenamefont {Cheong},
  \citenamefont {Nallaiyan}, \citenamefont {Spagna}, \citenamefont {Maple},
  \citenamefont {Nevidomskyy}, \citenamefont {Morosan}, \citenamefont {Chen},\
  and\ \citenamefont
  {Dai}}]{Gao.Dai_NP19_ExperimentalSignaturesThreedimensionalb}%
  \BibitemOpen
  \bibfield  {author} {\bibinfo {author} {\bibfnamefont {B.}~\bibnamefont
  {Gao}}, \bibinfo {author} {\bibfnamefont {T.}~\bibnamefont {Chen}}, \bibinfo
  {author} {\bibfnamefont {D.~W.}\ \bibnamefont {Tam}}, \bibinfo {author}
  {\bibfnamefont {C.-L.}\ \bibnamefont {Huang}}, \bibinfo {author}
  {\bibfnamefont {K.}~\bibnamefont {Sasmal}}, \bibinfo {author} {\bibfnamefont
  {D.~T.}\ \bibnamefont {Adroja}}, \bibinfo {author} {\bibfnamefont
  {F.}~\bibnamefont {Ye}}, \bibinfo {author} {\bibfnamefont {H.}~\bibnamefont
  {Cao}}, \bibinfo {author} {\bibfnamefont {G.}~\bibnamefont {Sala}}, \bibinfo
  {author} {\bibfnamefont {M.~B.}\ \bibnamefont {Stone}}, \bibinfo {author}
  {\bibfnamefont {C.}~\bibnamefont {Baines}}, \bibinfo {author} {\bibfnamefont
  {J.~A.~T.}\ \bibnamefont {Verezhak}}, \bibinfo {author} {\bibfnamefont
  {H.}~\bibnamefont {Hu}}, \bibinfo {author} {\bibfnamefont {J.-H.}\
  \bibnamefont {Chung}}, \bibinfo {author} {\bibfnamefont {X.}~\bibnamefont
  {Xu}}, \bibinfo {author} {\bibfnamefont {S.-W.}\ \bibnamefont {Cheong}},
  \bibinfo {author} {\bibfnamefont {M.}~\bibnamefont {Nallaiyan}}, \bibinfo
  {author} {\bibfnamefont {S.}~\bibnamefont {Spagna}}, \bibinfo {author}
  {\bibfnamefont {M.~B.}\ \bibnamefont {Maple}}, \bibinfo {author}
  {\bibfnamefont {A.~H.}\ \bibnamefont {Nevidomskyy}}, \bibinfo {author}
  {\bibfnamefont {E.}~\bibnamefont {Morosan}}, \bibinfo {author} {\bibfnamefont
  {G.}~\bibnamefont {Chen}},\ and\ \bibinfo {author} {\bibfnamefont
  {P.}~\bibnamefont {Dai}},\ }\href {https://doi.org/10.1038/s41567-019-0577-6}
  {\bibfield  {journal} {\bibinfo  {journal} {Nature Physics}\ }\textbf
  {\bibinfo {volume} {15}},\ \bibinfo {pages} {1052} (\bibinfo {year}
  {2019})}\BibitemShut {NoStop}%
\bibitem [{\citenamefont {Baidya}\ \emph {et~al.}(2007)\citenamefont {Baidya},
  \citenamefont {Hegde},\ and\ \citenamefont
  {Gopalakrishnan}}]{Baidya.Gopalakrishnan_JPCB07_OxygenReleaseStoragePropertiesa}%
  \BibitemOpen
  \bibfield  {author} {\bibinfo {author} {\bibfnamefont {T.}~\bibnamefont
  {Baidya}}, \bibinfo {author} {\bibfnamefont {M.~S.}\ \bibnamefont {Hegde}},\
  and\ \bibinfo {author} {\bibfnamefont {J.}~\bibnamefont {Gopalakrishnan}},\
  }\href {https://doi.org/10.1021/jp070525e} {\bibfield  {journal} {\bibinfo
  {journal} {The Journal of Physical Chemistry B}\ }\textbf {\bibinfo {volume}
  {111}},\ \bibinfo {pages} {5149} (\bibinfo {year} {2007})}\BibitemShut
  {NoStop}%
\bibitem [{\citenamefont {Sibille}\ \emph {et~al.}(2020)\citenamefont
  {Sibille}, \citenamefont {Gauthier}, \citenamefont {Lhotel}, \citenamefont
  {Porée}, \citenamefont {Pomjakushin}, \citenamefont {Ewings}, \citenamefont
  {Perring}, \citenamefont {Ollivier}, \citenamefont {Wildes}, \citenamefont
  {Ritter}, \citenamefont {Hansen}, \citenamefont {Keen}, \citenamefont
  {Nilsen}, \citenamefont {Keller}, \citenamefont {Petit},\ and\ \citenamefont
  {Fennell}}]{Sibille.Fennell_NP20_QuantumLiquidMagneticb}%
  \BibitemOpen
  \bibfield  {author} {\bibinfo {author} {\bibfnamefont {R.}~\bibnamefont
  {Sibille}}, \bibinfo {author} {\bibfnamefont {N.}~\bibnamefont {Gauthier}},
  \bibinfo {author} {\bibfnamefont {E.}~\bibnamefont {Lhotel}}, \bibinfo
  {author} {\bibfnamefont {V.}~\bibnamefont {Porée}}, \bibinfo {author}
  {\bibfnamefont {V.}~\bibnamefont {Pomjakushin}}, \bibinfo {author}
  {\bibfnamefont {R.~A.}\ \bibnamefont {Ewings}}, \bibinfo {author}
  {\bibfnamefont {T.~G.}\ \bibnamefont {Perring}}, \bibinfo {author}
  {\bibfnamefont {J.}~\bibnamefont {Ollivier}}, \bibinfo {author}
  {\bibfnamefont {A.}~\bibnamefont {Wildes}}, \bibinfo {author} {\bibfnamefont
  {C.}~\bibnamefont {Ritter}}, \bibinfo {author} {\bibfnamefont {T.~C.}\
  \bibnamefont {Hansen}}, \bibinfo {author} {\bibfnamefont {D.~A.}\
  \bibnamefont {Keen}}, \bibinfo {author} {\bibfnamefont {G.~J.}\ \bibnamefont
  {Nilsen}}, \bibinfo {author} {\bibfnamefont {L.}~\bibnamefont {Keller}},
  \bibinfo {author} {\bibfnamefont {S.}~\bibnamefont {Petit}},\ and\ \bibinfo
  {author} {\bibfnamefont {T.}~\bibnamefont {Fennell}},\ }\href
  {https://doi.org/10.1038/s41567-020-0827-7} {\bibfield  {journal} {\bibinfo
  {journal} {Nature Physics}\ }\textbf {\bibinfo {volume} {16}},\ \bibinfo
  {pages} {546} (\bibinfo {year} {2020})},\ \bibinfo {note} {arXiv:1912.00928
  [cond-mat.str-el]}\BibitemShut {NoStop}%
\bibitem [{\citenamefont {Smith}\ \emph {et~al.}(2022)\citenamefont {Smith},
  \citenamefont {Benton}, \citenamefont {Yahne}, \citenamefont {Placke},
  \citenamefont {Schäfer}, \citenamefont {Gaudet}, \citenamefont {Dudemaine},
  \citenamefont {Fitterman}, \citenamefont {Beare}, \citenamefont {Wildes},
  \citenamefont {Bhattacharya}, \citenamefont {DeLazzer}, \citenamefont
  {Buhariwalla}, \citenamefont {Butch}, \citenamefont {Movshovich},
  \citenamefont {Garrett}, \citenamefont {Marjerrison}, \citenamefont {Clancy},
  \citenamefont {Kermarrec}, \citenamefont {Luke}, \citenamefont {Bianchi},
  \citenamefont {Ross},\ and\ \citenamefont
  {Gaulin}}]{Smith.Gaulin_PRX22_Case$mathrmU1_ensuremathpi$Quantuma}%
  \BibitemOpen
  \bibfield  {author} {\bibinfo {author} {\bibfnamefont {E.}~\bibnamefont
  {Smith}}, \bibinfo {author} {\bibfnamefont {O.}~\bibnamefont {Benton}},
  \bibinfo {author} {\bibfnamefont {D.}~\bibnamefont {Yahne}}, \bibinfo
  {author} {\bibfnamefont {B.}~\bibnamefont {Placke}}, \bibinfo {author}
  {\bibfnamefont {R.}~\bibnamefont {Schäfer}}, \bibinfo {author}
  {\bibfnamefont {J.}~\bibnamefont {Gaudet}}, \bibinfo {author} {\bibfnamefont
  {J.}~\bibnamefont {Dudemaine}}, \bibinfo {author} {\bibfnamefont
  {A.}~\bibnamefont {Fitterman}}, \bibinfo {author} {\bibfnamefont
  {J.}~\bibnamefont {Beare}}, \bibinfo {author} {\bibfnamefont
  {A.}~\bibnamefont {Wildes}}, \bibinfo {author} {\bibfnamefont
  {S.}~\bibnamefont {Bhattacharya}}, \bibinfo {author} {\bibfnamefont
  {T.}~\bibnamefont {DeLazzer}}, \bibinfo {author} {\bibfnamefont
  {C.}~\bibnamefont {Buhariwalla}}, \bibinfo {author} {\bibfnamefont
  {N.}~\bibnamefont {Butch}}, \bibinfo {author} {\bibfnamefont
  {R.}~\bibnamefont {Movshovich}}, \bibinfo {author} {\bibfnamefont
  {J.}~\bibnamefont {Garrett}}, \bibinfo {author} {\bibfnamefont
  {C.}~\bibnamefont {Marjerrison}}, \bibinfo {author} {\bibfnamefont
  {J.}~\bibnamefont {Clancy}}, \bibinfo {author} {\bibfnamefont
  {E.}~\bibnamefont {Kermarrec}}, \bibinfo {author} {\bibfnamefont
  {G.}~\bibnamefont {Luke}}, \bibinfo {author} {\bibfnamefont {A.}~\bibnamefont
  {Bianchi}}, \bibinfo {author} {\bibfnamefont {K.}~\bibnamefont {Ross}},\ and\
  \bibinfo {author} {\bibfnamefont {B.}~\bibnamefont {Gaulin}},\ }\href
  {https://doi.org/10.1103/PhysRevX.12.021015} {\bibfield  {journal} {\bibinfo
  {journal} {Physical Review X}\ }\textbf {\bibinfo {volume} {12}},\ \bibinfo
  {pages} {021015} (\bibinfo {year} {2022})}\BibitemShut {NoStop}%
\bibitem [{\citenamefont {Porée}\ \emph {et~al.}(2022)\citenamefont {Porée},
  \citenamefont {Lhotel}, \citenamefont {Petit}, \citenamefont {Krajewska},
  \citenamefont {Puphal}, \citenamefont {Clark}, \citenamefont {Pomjakushin},
  \citenamefont {Walker}, \citenamefont {Gauthier}, \citenamefont {Gawryluk},\
  and\ \citenamefont
  {Sibille}}]{Poree.Sibille_PRM22_CrystalfieldStatesDefectb}%
  \BibitemOpen
  \bibfield  {author} {\bibinfo {author} {\bibfnamefont {V.}~\bibnamefont
  {Porée}}, \bibinfo {author} {\bibfnamefont {E.}~\bibnamefont {Lhotel}},
  \bibinfo {author} {\bibfnamefont {S.}~\bibnamefont {Petit}}, \bibinfo
  {author} {\bibfnamefont {A.}~\bibnamefont {Krajewska}}, \bibinfo {author}
  {\bibfnamefont {P.}~\bibnamefont {Puphal}}, \bibinfo {author} {\bibfnamefont
  {A.~H.}\ \bibnamefont {Clark}}, \bibinfo {author} {\bibfnamefont
  {V.}~\bibnamefont {Pomjakushin}}, \bibinfo {author} {\bibfnamefont {H.~C.}\
  \bibnamefont {Walker}}, \bibinfo {author} {\bibfnamefont {N.}~\bibnamefont
  {Gauthier}}, \bibinfo {author} {\bibfnamefont {D.~J.}\ \bibnamefont
  {Gawryluk}},\ and\ \bibinfo {author} {\bibfnamefont {R.}~\bibnamefont
  {Sibille}},\ }\href {https://doi.org/10.1103/PhysRevMaterials.6.044406}
  {\bibfield  {journal} {\bibinfo  {journal} {Physical Review Materials}\
  }\textbf {\bibinfo {volume} {6}},\ \bibinfo {pages} {044406} (\bibinfo {year}
  {2022})}\BibitemShut {NoStop}%
\bibitem [{\citenamefont {Yahne}\ \emph {et~al.}(2024)\citenamefont {Yahne},
  \citenamefont {Placke}, \citenamefont {Schäfer}, \citenamefont {Benton},
  \citenamefont {Moessner}, \citenamefont {Powell}, \citenamefont {Kolis},
  \citenamefont {Pasco}, \citenamefont {May}, \citenamefont {Frontzek},
  \citenamefont {Smith}, \citenamefont {Gaulin}, \citenamefont {Calder},\ and\
  \citenamefont {Ross}}]{Yahne.Ross_PRX24_DipolarSpinIce}%
  \BibitemOpen
  \bibfield  {author} {\bibinfo {author} {\bibfnamefont {D.~R.}\ \bibnamefont
  {Yahne}}, \bibinfo {author} {\bibfnamefont {B.}~\bibnamefont {Placke}},
  \bibinfo {author} {\bibfnamefont {R.}~\bibnamefont {Schäfer}}, \bibinfo
  {author} {\bibfnamefont {O.}~\bibnamefont {Benton}}, \bibinfo {author}
  {\bibfnamefont {R.}~\bibnamefont {Moessner}}, \bibinfo {author}
  {\bibfnamefont {M.}~\bibnamefont {Powell}}, \bibinfo {author} {\bibfnamefont
  {J.~W.}\ \bibnamefont {Kolis}}, \bibinfo {author} {\bibfnamefont {C.~M.}\
  \bibnamefont {Pasco}}, \bibinfo {author} {\bibfnamefont {A.~F.}\ \bibnamefont
  {May}}, \bibinfo {author} {\bibfnamefont {M.~D.}\ \bibnamefont {Frontzek}},
  \bibinfo {author} {\bibfnamefont {E.~M.}\ \bibnamefont {Smith}}, \bibinfo
  {author} {\bibfnamefont {B.~D.}\ \bibnamefont {Gaulin}}, \bibinfo {author}
  {\bibfnamefont {S.}~\bibnamefont {Calder}},\ and\ \bibinfo {author}
  {\bibfnamefont {K.~A.}\ \bibnamefont {Ross}},\ }\href
  {https://doi.org/10.1103/PhysRevX.14.011005} {\bibfield  {journal} {\bibinfo
  {journal} {Physical Review X}\ }\textbf {\bibinfo {volume} {14}},\ \bibinfo
  {pages} {011005} (\bibinfo {year} {2024})}\BibitemShut {NoStop}%
\bibitem [{\citenamefont {Smith}\ \emph {et~al.}(2023)\citenamefont {Smith},
  \citenamefont {Dudemaine}, \citenamefont {Placke}, \citenamefont {Schäfer},
  \citenamefont {Yahne}, \citenamefont {DeLazzer}, \citenamefont {Fitterman},
  \citenamefont {Beare}, \citenamefont {Gaudet}, \citenamefont {Buhariwalla},
  \citenamefont {Podlesnyak}, \citenamefont {Xu}, \citenamefont {Clancy},
  \citenamefont {Movshovich}, \citenamefont {Luke}, \citenamefont {Ross},
  \citenamefont {Moessner}, \citenamefont {Benton}, \citenamefont {Bianchi},\
  and\ \citenamefont {Gaulin}}]{Smith.Gaulin_PRB23_QuantumSpinIceb}%
  \BibitemOpen
  \bibfield  {author} {\bibinfo {author} {\bibfnamefont {E.~M.}\ \bibnamefont
  {Smith}}, \bibinfo {author} {\bibfnamefont {J.}~\bibnamefont {Dudemaine}},
  \bibinfo {author} {\bibfnamefont {B.}~\bibnamefont {Placke}}, \bibinfo
  {author} {\bibfnamefont {R.}~\bibnamefont {Schäfer}}, \bibinfo {author}
  {\bibfnamefont {D.~R.}\ \bibnamefont {Yahne}}, \bibinfo {author}
  {\bibfnamefont {T.}~\bibnamefont {DeLazzer}}, \bibinfo {author}
  {\bibfnamefont {A.}~\bibnamefont {Fitterman}}, \bibinfo {author}
  {\bibfnamefont {J.}~\bibnamefont {Beare}}, \bibinfo {author} {\bibfnamefont
  {J.}~\bibnamefont {Gaudet}}, \bibinfo {author} {\bibfnamefont {C.~R.~C.}\
  \bibnamefont {Buhariwalla}}, \bibinfo {author} {\bibfnamefont
  {A.}~\bibnamefont {Podlesnyak}}, \bibinfo {author} {\bibfnamefont
  {G.}~\bibnamefont {Xu}}, \bibinfo {author} {\bibfnamefont {J.~P.}\
  \bibnamefont {Clancy}}, \bibinfo {author} {\bibfnamefont {R.}~\bibnamefont
  {Movshovich}}, \bibinfo {author} {\bibfnamefont {G.~M.}\ \bibnamefont
  {Luke}}, \bibinfo {author} {\bibfnamefont {K.~A.}\ \bibnamefont {Ross}},
  \bibinfo {author} {\bibfnamefont {R.}~\bibnamefont {Moessner}}, \bibinfo
  {author} {\bibfnamefont {O.}~\bibnamefont {Benton}}, \bibinfo {author}
  {\bibfnamefont {A.~D.}\ \bibnamefont {Bianchi}},\ and\ \bibinfo {author}
  {\bibfnamefont {B.~D.}\ \bibnamefont {Gaulin}},\ }\href
  {https://doi.org/10.1103/PhysRevB.108.054438} {\bibfield  {journal} {\bibinfo
   {journal} {Physical Review B}\ }\textbf {\bibinfo {volume} {108}},\ \bibinfo
  {pages} {054438} (\bibinfo {year} {2023})}\BibitemShut {NoStop}%
\bibitem [{\citenamefont {Gao}\ \emph {et~al.}(2025)\citenamefont {Gao},
  \citenamefont {Desrochers}, \citenamefont {Tam}, \citenamefont {Kirschbaum},
  \citenamefont {Steffens}, \citenamefont {Hiess}, \citenamefont {Nguyen},
  \citenamefont {Su}, \citenamefont {Cheong}, \citenamefont {Paschen},
  \citenamefont {Kim},\ and\ \citenamefont
  {Dai}}]{Gao.Dai_NP25_NeutronScatteringThermodynamica}%
  \BibitemOpen
  \bibfield  {author} {\bibinfo {author} {\bibfnamefont {B.}~\bibnamefont
  {Gao}}, \bibinfo {author} {\bibfnamefont {F.}~\bibnamefont {Desrochers}},
  \bibinfo {author} {\bibfnamefont {D.~W.}\ \bibnamefont {Tam}}, \bibinfo
  {author} {\bibfnamefont {D.~M.}\ \bibnamefont {Kirschbaum}}, \bibinfo
  {author} {\bibfnamefont {P.}~\bibnamefont {Steffens}}, \bibinfo {author}
  {\bibfnamefont {A.}~\bibnamefont {Hiess}}, \bibinfo {author} {\bibfnamefont
  {D.~H.}\ \bibnamefont {Nguyen}}, \bibinfo {author} {\bibfnamefont
  {Y.}~\bibnamefont {Su}}, \bibinfo {author} {\bibfnamefont {S.-W.}\
  \bibnamefont {Cheong}}, \bibinfo {author} {\bibfnamefont {S.}~\bibnamefont
  {Paschen}}, \bibinfo {author} {\bibfnamefont {Y.~B.}\ \bibnamefont {Kim}},\
  and\ \bibinfo {author} {\bibfnamefont {P.}~\bibnamefont {Dai}},\ }\href
  {https://doi.org/10.1038/s41567-025-02922-9} {\bibfield  {journal} {\bibinfo
  {journal} {Nature Physics}\ }\textbf {\bibinfo {volume} {21}},\ \bibinfo
  {pages} {1203} (\bibinfo {year} {2025})}\BibitemShut {NoStop}%
\bibitem [{\citenamefont {Huang}\ \emph {et~al.}(2014)\citenamefont {Huang},
  \citenamefont {Chen},\ and\ \citenamefont
  {Hermele}}]{Huang.Hermele_PRL14_QuantumSpinIcesb}%
  \BibitemOpen
  \bibfield  {author} {\bibinfo {author} {\bibfnamefont {Y.-P.}\ \bibnamefont
  {Huang}}, \bibinfo {author} {\bibfnamefont {G.}~\bibnamefont {Chen}},\ and\
  \bibinfo {author} {\bibfnamefont {M.}~\bibnamefont {Hermele}},\ }\href
  {https://doi.org/10.1103/PhysRevLett.112.167203} {\bibfield  {journal}
  {\bibinfo  {journal} {Physical Review Letters}\ }\textbf {\bibinfo {volume}
  {112}},\ \bibinfo {pages} {167203} (\bibinfo {year} {2014})}\BibitemShut
  {NoStop}%
\bibitem [{\citenamefont {Li}\ and\ \citenamefont
  {Chen}(2017)}]{Li.Chen_PRB17_SymmetryEnrichedU1b}%
  \BibitemOpen
  \bibfield  {author} {\bibinfo {author} {\bibfnamefont {Y.-D.}\ \bibnamefont
  {Li}}\ and\ \bibinfo {author} {\bibfnamefont {G.}~\bibnamefont {Chen}},\
  }\href {https://doi.org/10.1103/PhysRevB.95.041106} {\bibfield  {journal}
  {\bibinfo  {journal} {Physical Review B}\ }\textbf {\bibinfo {volume} {95}},\
  \bibinfo {pages} {041106} (\bibinfo {year} {2017})}\BibitemShut {NoStop}%
\bibitem [{\citenamefont
  {Benton}(2020)}]{Benton_PRB20_GroundstatePhaseDiagramb}%
  \BibitemOpen
  \bibfield  {author} {\bibinfo {author} {\bibfnamefont {O.}~\bibnamefont
  {Benton}},\ }\href {https://doi.org/10.1103/PhysRevB.102.104408} {\bibfield
  {journal} {\bibinfo  {journal} {Physical Review B}\ }\textbf {\bibinfo
  {volume} {102}},\ \bibinfo {pages} {104408} (\bibinfo {year}
  {2020})}\BibitemShut {NoStop}%
\bibitem [{\citenamefont {Bhardwaj}\ \emph {et~al.}(2022)\citenamefont
  {Bhardwaj}, \citenamefont {Zhang}, \citenamefont {Yan}, \citenamefont
  {Moessner}, \citenamefont {Nevidomskyy},\ and\ \citenamefont
  {Changlani}}]{Bhardwaj.Changlani_nQM22_SleuthingOutExotic}%
  \BibitemOpen
  \bibfield  {author} {\bibinfo {author} {\bibfnamefont {A.}~\bibnamefont
  {Bhardwaj}}, \bibinfo {author} {\bibfnamefont {S.}~\bibnamefont {Zhang}},
  \bibinfo {author} {\bibfnamefont {H.}~\bibnamefont {Yan}}, \bibinfo {author}
  {\bibfnamefont {R.}~\bibnamefont {Moessner}}, \bibinfo {author}
  {\bibfnamefont {A.~H.}\ \bibnamefont {Nevidomskyy}},\ and\ \bibinfo {author}
  {\bibfnamefont {H.~J.}\ \bibnamefont {Changlani}},\ }\href
  {https://doi.org/10.1038/s41535-022-00458-2} {\bibfield  {journal} {\bibinfo
  {journal} {npj Quantum Materials}\ }\textbf {\bibinfo {volume} {7}},\
  \bibinfo {pages} {51} (\bibinfo {year} {2022})}\BibitemShut {NoStop}%
\bibitem [{\citenamefont {Desrochers}\ \emph {et~al.}(2023)\citenamefont
  {Desrochers}, \citenamefont {Chern},\ and\ \citenamefont
  {Kim}}]{Desrochers.Kim_PRB23_SymmetryFractionalizationGaugeb}%
  \BibitemOpen
  \bibfield  {author} {\bibinfo {author} {\bibfnamefont {F.}~\bibnamefont
  {Desrochers}}, \bibinfo {author} {\bibfnamefont {L.~E.}\ \bibnamefont
  {Chern}},\ and\ \bibinfo {author} {\bibfnamefont {Y.~B.}\ \bibnamefont
  {Kim}},\ }\href {https://doi.org/10.1103/PhysRevB.107.064404} {\bibfield
  {journal} {\bibinfo  {journal} {Physical Review B}\ }\textbf {\bibinfo
  {volume} {107}},\ \bibinfo {pages} {064404} (\bibinfo {year}
  {2023})}\BibitemShut {NoStop}%
\bibitem [{\citenamefont {Desrochers}\ and\ \citenamefont
  {Kim}(2024)}]{Desrochers.Kim_PRL24_SpectroscopicSignaturesFractionalizationb}%
  \BibitemOpen
  \bibfield  {author} {\bibinfo {author} {\bibfnamefont {F.}~\bibnamefont
  {Desrochers}}\ and\ \bibinfo {author} {\bibfnamefont {Y.~B.}\ \bibnamefont
  {Kim}},\ }\href {https://doi.org/10.1103/PhysRevLett.132.066502} {\bibfield
  {journal} {\bibinfo  {journal} {Physical Review Letters}\ }\textbf {\bibinfo
  {volume} {132}},\ \bibinfo {pages} {066502} (\bibinfo {year}
  {2024})}\BibitemShut {NoStop}%
\bibitem [{\citenamefont {Hosoi}\ \emph {et~al.}(2022)\citenamefont {Hosoi},
  \citenamefont {Zhang}, \citenamefont {Patri},\ and\ \citenamefont
  {Kim}}]{Hosoi.Kim_PRL22_UncoveringFootprintsDipolarOctupolarb}%
  \BibitemOpen
  \bibfield  {author} {\bibinfo {author} {\bibfnamefont {M.}~\bibnamefont
  {Hosoi}}, \bibinfo {author} {\bibfnamefont {E.~Z.}\ \bibnamefont {Zhang}},
  \bibinfo {author} {\bibfnamefont {A.~S.}\ \bibnamefont {Patri}},\ and\
  \bibinfo {author} {\bibfnamefont {Y.~B.}\ \bibnamefont {Kim}},\ }\href
  {https://doi.org/10.1103/PhysRevLett.129.097202} {\bibfield  {journal}
  {\bibinfo  {journal} {Physical Review Letters}\ }\textbf {\bibinfo {volume}
  {129}},\ \bibinfo {pages} {097202} (\bibinfo {year} {2022})}\BibitemShut
  {NoStop}%
\bibitem [{Note5()}]{Note5}%
  \BibitemOpen
  \bibinfo {note} {This is reasonable in the strong Ising limit of the QSI,
  where all these scales are controlled by the ring exchange \cite
  {Hermele.Balents_PRB04_PyrochlorePhotons$U1$}.}\BibitemShut {Stop}%
\bibitem [{\citenamefont {Uno}\ \emph {et~al.}(2006)\citenamefont {Uno},
  \citenamefont {Kosuga}, \citenamefont {Okui}, \citenamefont {Horisaka},
  \citenamefont {Muta}, \citenamefont {Kurosaki},\ and\ \citenamefont
  {Yamanaka}}]{Uno.Yamanaka_JAC06_PhotoelectrochemicalStudyLanthanide}%
  \BibitemOpen
  \bibfield  {author} {\bibinfo {author} {\bibfnamefont {M.}~\bibnamefont
  {Uno}}, \bibinfo {author} {\bibfnamefont {A.}~\bibnamefont {Kosuga}},
  \bibinfo {author} {\bibfnamefont {M.}~\bibnamefont {Okui}}, \bibinfo {author}
  {\bibfnamefont {K.}~\bibnamefont {Horisaka}}, \bibinfo {author}
  {\bibfnamefont {H.}~\bibnamefont {Muta}}, \bibinfo {author} {\bibfnamefont
  {K.}~\bibnamefont {Kurosaki}},\ and\ \bibinfo {author} {\bibfnamefont
  {S.}~\bibnamefont {Yamanaka}},\ }\href
  {https://doi.org/10.1016/j.jallcom.2005.10.072} {\bibfield  {journal}
  {\bibinfo  {journal} {Journal of Alloys and Compounds}\ }\textbf {\bibinfo
  {volume} {420}},\ \bibinfo {pages} {291} (\bibinfo {year}
  {2006})}\BibitemShut {NoStop}%
\bibitem [{\citenamefont {Wilson}(1974)}]{Wilson_PRD74_ConfinementQuarks}%
  \BibitemOpen
  \bibfield  {author} {\bibinfo {author} {\bibfnamefont {K.~G.}\ \bibnamefont
  {Wilson}},\ }\href {https://doi.org/10.1103/PhysRevD.10.2445} {\bibfield
  {journal} {\bibinfo  {journal} {Phys. Rev. D}\ }\textbf {\bibinfo {volume}
  {10}},\ \bibinfo {pages} {2445} (\bibinfo {year} {1974})}\BibitemShut
  {NoStop}%
\bibitem [{\citenamefont {Kogut}\ and\ \citenamefont
  {Susskind}(1975)}]{Kogut.Susskind_PRD75_HamiltonianFormulationWilsonsb}%
  \BibitemOpen
  \bibfield  {author} {\bibinfo {author} {\bibfnamefont {J.}~\bibnamefont
  {Kogut}}\ and\ \bibinfo {author} {\bibfnamefont {L.}~\bibnamefont
  {Susskind}},\ }\href {https://doi.org/10.1103/PhysRevD.11.395} {\bibfield
  {journal} {\bibinfo  {journal} {Phys. Rev. D}\ }\textbf {\bibinfo {volume}
  {11}},\ \bibinfo {pages} {395} (\bibinfo {year} {1975})}\BibitemShut
  {NoStop}%
\bibitem [{\citenamefont
  {Kogut}(1979)}]{Kogut_RMP79_IntroductionLatticeGaugea}%
  \BibitemOpen
  \bibfield  {author} {\bibinfo {author} {\bibfnamefont {J.~B.}\ \bibnamefont
  {Kogut}},\ }\href {https://doi.org/10.1103/RevModPhys.51.659} {\bibfield
  {journal} {\bibinfo  {journal} {Rev. Mod. Phys.}\ }\textbf {\bibinfo {volume}
  {51}},\ \bibinfo {pages} {659} (\bibinfo {year} {1979})}\BibitemShut
  {NoStop}%
\bibitem [{Note6()}]{Note6}%
  \BibitemOpen
  \bibinfo {note} {For microscopic calculations of quasiparticle properties of
  pyrochlore 3D $U(1)$ QSLs arising in QSI, see e.g. \cite
  {Hermele.Balents_PRB04_PyrochlorePhotons$U1$,
  Gingras.McClarty_RPP14_QuantumSpinIce, savary_coulombic_2012,
  Shannon.Fulde_PRL12_QuantumIceQuantum,
  Benton.Shannon_PRB12_SeeingLightExperimental,
  Lee.Balents_PRB12_GenericQuantumSpin}}\BibitemShut {NoStop}%
\bibitem [{Note7()}]{Note7}%
  \BibitemOpen
  \bibinfo {note} {The lattice curl is defined as $\protect \lcurl
  _{p}(a_l)\equiv \DOTSB \sum@ \slimits@ _{l\in p}\sigma _l^{p}a_{l}$ where
  $\sigma _l^p$ takes values $\pm 1$ along the links that bound the plaquette;
  see Appendix \ref {appendixA1} for precise definitions of lattice
  differential operators ($\protect \lgrad ,\protect \lcurl ,\protect \ldiv
  $).}\BibitemShut {Stop}%
\bibitem [{Note8()}]{Note8}%
  \BibitemOpen
  \bibinfo {note} {The theory has a single dimensionless parameter $\kappa
  =\hbar ^{-2}\epsilon /\mu $. For $\kappa \ll 1$ it is in the deconfined
  Maxwell phase (quantum spin-liquid), and transitions into the confined phase
  (trivial paramagnet) for $\kappa >\kappa _{\protect \rm crit} \sim
  1$.}\BibitemShut {Stop}%
\bibitem [{Note9()}]{Note9}%
  \BibitemOpen
  \bibinfo {note} {The discrete divergence on a vertex $v$ is $\protect \ldiv
  _v(e_l)=\DOTSB \sum@ \slimits@ _{l\in v}\sigma ^{v}_{l}e_l$, where $\sigma
  ^{v}_{l}$ takes values $\pm 1$. See Appendix \ref {appendixA1} for sign
  convention.}\BibitemShut {Stop}%
\bibitem [{Note10()}]{Note10}%
  \BibitemOpen
  \bibinfo {note} {The discrete divergence at a dual cube center $c$ is
  $\protect \ldiv _c(g_p)=\DOTSB \sum@ \slimits@ _{p\in c}\sigma _p^{c}g_{p}$,
  where $\sigma ^{p}_{c}$ takes values $\pm 1$. See Appendix \ref {appendixA1}
  for sign convention.}\BibitemShut {Stop}%
\bibitem [{Note11()}]{Note11}%
  \BibitemOpen
  \bibinfo {note} {Namely they are not viewed as dynamical fields.}\BibitemShut
  {Stop}%
\bibitem [{Note12()}]{Note12}%
  \BibitemOpen
  \bibinfo {note} {It is straightforward to generalize our analysis to the case
  in which several species of monopoles with different charges are thermally
  activated.}\BibitemShut {Stop}%
\bibitem [{Note13()}]{Note13}%
  \BibitemOpen
  \bibinfo {note} {Within linear response, all physical quantities, e.g.
  $J_\protect \text {m}$ and $g$, are spatially uniform and oscillate with the
  same frequency of the external perturbation $E$.}\BibitemShut {Stop}%
\bibitem [{Note14()}]{Note14}%
  \BibitemOpen
  \bibinfo {note} {Monopoles of opposite charges will have opposite velocities
  and their currents add up. Thus $n_\protect \text {m}(T)$ stands here for the
  net positive density of both positive and negative monopoles.}\BibitemShut
  {Stop}%
\bibitem [{Note15()}]{Note15}%
  \BibitemOpen
  \bibinfo {note} {It is interesting to note that we have arrived at this rule
  by a direct analysis of the dynamics of gauge fields and monopoles without
  resorting to parton constructions, which are more commonly used to derive
  it.}\BibitemShut {Stop}%
\bibitem [{\citenamefont {Oxenius}(1986)}]{Oxenius1986}%
  \BibitemOpen
  \bibfield  {author} {\bibinfo {author} {\bibfnamefont {J.}~\bibnamefont
  {Oxenius}},\ }\bibinfo {title} {Momentum exchange between matter and
  radiation},\ in\ \href {https://doi.org/10.1007/978-3-642-70728-5_7} {\emph
  {\bibinfo {booktitle} {Kinetic Theory of Particles and Photons: Theoretical
  Foundations of Non-LTE Plasma Spectroscopy}}}\ (\bibinfo  {publisher}
  {Springer Berlin Heidelberg},\ \bibinfo {address} {Berlin, Heidelberg},\
  \bibinfo {year} {1986})\ pp.\ \bibinfo {pages} {240--273}\BibitemShut
  {NoStop}%
\bibitem [{Note16()}]{Note16}%
  \BibitemOpen
  \bibinfo {note} {Notice that emergent electric associated with spin-ice
  violation charges are always neutral, e.g. by time-reversal or inversion
  symmetries.}\BibitemShut {Stop}%
\bibitem [{Note17()}]{Note17}%
  \BibitemOpen
  \bibinfo {note} {This is already achieved for Ce$_2$Zr$_2$O$_7$ in e.g.
  Ref.\cite {Gao.Dai_NP25_NeutronScatteringThermodynamica}, since at their
  lowest temperatures the specific heat is dominated by emergent photons. In
  optical experiments it will be important to monitor that there is no light
  induced heating that could substantially raise the temperature of
  spins.}\BibitemShut {Stop}%
\bibitem [{Note18()}]{Note18}%
  \BibitemOpen
  \bibinfo {note} {If instead we assume that {\protect \it all} the monopole
  energy scales are comparable to the photon bandwidth, the conductivity
  changes with temperature in an almost similar fashion at different driving
  frequencies (see Appendix \ref {ap:regimes}). The conductivity peaks at
  $T^\ast \approx 25\protect \,\protect \text {mK}$, and the maximum value at
  the peak becomes $\sigma _\protect \text {max}\approx 0.005\protect
  \,e_\protect \text {m}^2/(ah)$, which is somewhat smaller, but still
  sizable.}\BibitemShut {Stop}%
\bibitem [{\citenamefont {Savary}\ and\ \citenamefont
  {Balents}(2012)}]{savary_coulombic_2012}%
  \BibitemOpen
  \bibfield  {author} {\bibinfo {author} {\bibfnamefont {L.}~\bibnamefont
  {Savary}}\ and\ \bibinfo {author} {\bibfnamefont {L.}~\bibnamefont
  {Balents}},\ }\href {https://doi.org/10.1103/PhysRevLett.108.037202}
  {\bibfield  {journal} {\bibinfo  {journal} {Physical Review Letters}\
  }\textbf {\bibinfo {volume} {108}},\ \bibinfo {pages} {037202} (\bibinfo
  {year} {2012})}\BibitemShut {NoStop}%
\bibitem [{\citenamefont {Shannon}\ \emph {et~al.}(2012)\citenamefont
  {Shannon}, \citenamefont {Sikora}, \citenamefont {Pollmann}, \citenamefont
  {Penc},\ and\ \citenamefont {Fulde}}]{Shannon.Fulde_PRL12_QuantumIceQuantum}%
  \BibitemOpen
  \bibfield  {author} {\bibinfo {author} {\bibfnamefont {N.}~\bibnamefont
  {Shannon}}, \bibinfo {author} {\bibfnamefont {O.}~\bibnamefont {Sikora}},
  \bibinfo {author} {\bibfnamefont {F.}~\bibnamefont {Pollmann}}, \bibinfo
  {author} {\bibfnamefont {K.}~\bibnamefont {Penc}},\ and\ \bibinfo {author}
  {\bibfnamefont {P.}~\bibnamefont {Fulde}},\ }\href
  {https://doi.org/10.1103/PhysRevLett.108.067204} {\bibfield  {journal}
  {\bibinfo  {journal} {Physical Review Letters}\ }\textbf {\bibinfo {volume}
  {108}},\ \bibinfo {pages} {067204} (\bibinfo {year} {2012})}\BibitemShut
  {NoStop}%
\bibitem [{\citenamefont {Benton}\ \emph {et~al.}(2012)\citenamefont {Benton},
  \citenamefont {Sikora},\ and\ \citenamefont
  {Shannon}}]{Benton.Shannon_PRB12_SeeingLightExperimental}%
  \BibitemOpen
  \bibfield  {author} {\bibinfo {author} {\bibfnamefont {O.}~\bibnamefont
  {Benton}}, \bibinfo {author} {\bibfnamefont {O.}~\bibnamefont {Sikora}},\
  and\ \bibinfo {author} {\bibfnamefont {N.}~\bibnamefont {Shannon}},\ }\href
  {https://doi.org/10.1103/PhysRevB.86.075154} {\bibfield  {journal} {\bibinfo
  {journal} {Physical Review B}\ }\textbf {\bibinfo {volume} {86}},\ \bibinfo
  {pages} {075154} (\bibinfo {year} {2012})}\BibitemShut {NoStop}%
\bibitem [{\citenamefont {Lee}\ \emph {et~al.}(2012)\citenamefont {Lee},
  \citenamefont {Onoda},\ and\ \citenamefont
  {Balents}}]{Lee.Balents_PRB12_GenericQuantumSpin}%
  \BibitemOpen
  \bibfield  {author} {\bibinfo {author} {\bibfnamefont {S.}~\bibnamefont
  {Lee}}, \bibinfo {author} {\bibfnamefont {S.}~\bibnamefont {Onoda}},\ and\
  \bibinfo {author} {\bibfnamefont {L.}~\bibnamefont {Balents}},\ }\href
  {https://doi.org/10.1103/PhysRevB.86.104412} {\bibfield  {journal} {\bibinfo
  {journal} {Phys. Rev. B}\ }\textbf {\bibinfo {volume} {86}},\ \bibinfo
  {pages} {104412} (\bibinfo {year} {2012})}\BibitemShut {NoStop}%
\end{thebibliography}%
\clearpage

\onecolumngrid

\section*{
--- Supplemental Material --- \\*[1em]
The monopole plasma resonance: a smoking gun of 3D $U(1)$ quantum spin liquids}

\beginsupplement
%%%%%%%%%%%%%%%%%%%%%%%%%%%%%%%%%%%%%%%%%%%%%%%%%%%%%%%%%%%%%%%%%%%%%%%%%%%%%%%%%%%%%
\section{Derivation of modified Maxwell's equations}
\label{appendixA1}
\begin{figure}[h]
\centering
% Equation on the left
\begin{minipage}{0.45\columnwidth}
\centering
\[
\frac{\partial b_p}{\partial a_l}=
\begin{cases}
-1 & \text{Plaquette 1}\\
-1 & \text{Plaquette 2}\\
+1 & \text{Plaquette 3}\\
+1 & \text{Plaquette 4}
\end{cases}
\]
\end{minipage}
\hspace{0.02\columnwidth}
% Figure on the right
\begin{minipage}{0.35\columnwidth}
\centering
\includegraphics[width=4cm]{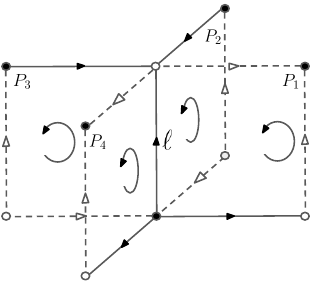}
\caption{Illustration of lattice curl of a link.}
\label{Fig4}
\end{minipage}

\end{figure}

Before we start with our Lagrangian to derive the modified Maxwell's equation, we first define the lattice derivatives as follows.
\begin{align}
    \begin{aligned}
        \ldiv_v^\bullet Q=\sum_{i\in v^\bullet}\sigma^v_i Q_i,
        \quad
        \lcurl_p(P_l)=\sum_{l\in p }\sigma_l^p P_l.
    \end{aligned}
\end{align}
Here Q and P are any quantity whose $\ldiv$ or $\lcurl$ is needed. For example, if $Q_i=\dot{a}_l$, we then have,
\begin{align}
    \ldiv_v^\bullet \dot{a}_l=\dot{a}_1+\dot{a}_2+\dot{a}_3-\dot{a}_4-\dot{a}_5-\dot{a}_6. \nonumber
\end{align}
In the above example, if we follow the 3-in 3-out ice rule, we get back $Q_e(v)=0$.
Note that in our system, we have an AB sub-lattice structure which translates to a N\'eel-like structure of $v^\bullet$ and $v^\circ$.
$\ldiv_v^\circ Q$ is related to $\ldiv_v^\bullet Q$ by a $-$ve sign i.e. $\ldiv_v^\circ Q=-\sum_{j\in v\circ} \sigma ^{v}_{j}Q_j$.
This ensures that if we a violate 1 bond from the ice rule, then we maintain the global emergent charge conservation by having $Q_e(v)=+2$ on vertex $v^\bullet_1$ and $Q_e(v)=-2$ on $v^\circ_2$.\\
For the case of $\lcurl_p(P_l)$, we decide the value of $\sigma_l^p$ using right-handed loop. For example, if $P_l=a_l$, we then have,
\begin{align}
   b_p\equiv \lcurl_p(a_l)&=a_1+a_2-a_3-a_4. \nonumber\\
\end{align}
The Lagrangian is written as,
\begin{align}
    \lagr=\sum_l\frac{\epsilon\dot{a}_l^2}{2}-\sum_p\frac{1-\cos(b_p)}{\mu}-\sum_lg_{eB}\dot{a}_lB_l(t)+\sum_p g_{bE}\sin(b_p)E_p(t).
\end{align}
We use the Euler-Lagrange equations of motion as follows,
\begin{align}
    \begin{aligned}
        \frac{\delta \lagr}{\delta \dot{a}_l}=\epsilon\dot{a}_l-g_{eB}B_l,
        \quad
        \quad
        \frac{\delta\lagr}{\delta a_l}=-\frac{\delta b_p}{\delta a_l}\biggl[\frac{1}{\mu}\sin(b_p)-g_{bE}\cos(b_p)E_p(t) \biggl].
    \end{aligned}
\end{align}
In order to calculate the quantity $\delta b_p/\delta a_l$, we refer to Fig.(\ref{Fig4}). Here we see only link $l$ is shared by all the 4 plaquettes.
Consider plaquette $P_1$ whose $b_p$ can be written as, $b_{p_1}=a_1+a_2-a_3-a_l$. 
Taking the derivative leads to the value of $-1$ for this plaquette.
The values of for the other plaquettes are summarized below.
\\
Because of this structure of $\delta b_p/\delta a_l$, we define it as the lattice curl of a link $\lcurl_l$.
The Euler-Lagrange equation of finally stands as,
\begin{align}
    \frac{d}{dt}\bigl[-\epsilon e_l-g_{eB}B_l\bigl]=-\lcurl_l\biggl[\frac{1}{\mu}\sin(b_p)-g_{bE}E_p\biggl].
\end{align}
Here we use $e_l=-da_l/dt$.
After linearization and coarse graining, we get the modified Ampere-Maxwell Eq.[\ref{Amp Maxwell's law}].
Taking a time derivative on the definition of $b_p$, we recover Eq.[\ref{Faraday's law}] in the following way,
\begin{align}
    d_t\bigl(g_p+2\pi M_p \bigl)=\lcurl_p\dot{a}_l \nonumber \\
    -[\dot{g}+2\pi\frac{dM_p}{dt}]=\lcurl_p e_l
\end{align}
The canonical momentum, $p_l$ given by $\delta \lagr/\delta \dot{a}_l$ leads us to the Electric Gauss law.
Since we are in the $Q_e(v)=0$ sector or equivalently $\ldiv_v p_l=0$, performing a spatial derivative on $p_l$ gives us,
\begin{align}
    \epsilon \ldiv_v e_l +g_{eB}\ldiv_v B_l = 0 
\end{align}
Coarse-graining the above equation leads us to Eq.[\ref{Gauss law}]. \\
In order to derive the Magnetic Gauss law, we need to prove $\ldiv_c b_p=0$.
 We define the dual divergence on the center of the cubic lattice (dual lattice) as follows, 
\begin{align}
    \ldiv_c f &=\sum_{p \in C} \tilde{\sigma}_p^{c} f_p \label{Defination of div_c} \\
            &= f_p^{+x} +f_p^{+y}+f_p^{+z}-f_p^{-x}-f_p^{-y}-f_p^{-z} \nonumber
\end{align}
For the case of $\ldiv_cb_p$, we replace $f_p$ with $\lcurl_pa_l$ where the dual cube comprises of 6 plaquettes.
\begin{align}
div_c b_p=(\underbrace{a_1+\cancel{a_2}-a_3-\bcancel{a_4}}_{f_p^{+x}})+(\underbrace{-a_5+a_6+a_7-\cancel{a_2}}_{f_p^{+y}})-(\underbrace{-a_8+a_9+a_{10}-\bcancel{a_4}}_{f_p^{-y}})+\underbrace{\hdots\hdots\hdots}_{f_p^{+z}, f_p^{-y}, f_p^{-z}}
\end{align}
The $a_l$'s cancel pairwise and we get $\ldiv_c b_p=0$.
Expressing $b_p=g_p+2\pi M_p$, we get Eq.[\ref{Magnetic Gauss law}].\\
%%%%%%%%%%%%%%%%%%%%%%%%%%%%%%%%%%%%%%%%%%%%%%%%%%%%%%%%%%%%%%%%%%%%%%%%%%%%%%%%%%%%%
\section{Derivation of Force on Monopole}
\label{appendixA2}
In order to derive the Newtonian force the monopole experiences, we will be re-expressing the $E_p(t)$ in $\lagr_{bE}$ as,
\begin{align}
   \sum_pg_{bE}\sin(b_p)\biggl[\underbracket{-\text{grad}_p\Phi_c(t)}_{\lagr_{bE}^{\Phi}}-\underbracket{\frac{dA_p(t)}{dt}}_{\lagr_{bE}^{A}} \biggl] \label{L coupling of b and E}.
\end{align}
In terms of lattice derivatives, we define $\text{grad}_p\Phi=\Phi_{c_2}-\Phi_{c_1}$ where $C_1$ and $C_2$ are dual centers of two consecutive cubes.
Using $b_p=g_p+2\pi$ and linearizing $\sin(g_p)$ to $g_p$,
\begin{align}
    \lagr_{bE}^{\Phi}&=
    -\sum_p g_{bE}g_p\text{grad}_p\bigl(\Phi_c(t)\bigl) 
\nonumber \\
   &=\sum_c g_{bE}\ldiv_c (g_p)\Phi_c(t) 
\nonumber \\
   &=\sum_c g_{bE}Q_\m(c)\Phi_c(t)
\end{align}
In the above derivation we made use of the fact $-\sum_c \ldiv_c(f_p)\Phi_c=\sum_p f_p\text{grad}_p(\Phi_c)$. To see this, consider two adjacent cubes and Eq.\eqref{Defination of div_c}.
\begin{align}
    \sum_c \ldiv_c(f_p)\Phi_c= \bigl( f_p^{+x} + f_p^{+y} + f_p^{+z}-f_p^{-x}-f_p^{-y}-f_p^{-z}\bigl)\Phi_{c_1}+\sum_{c\neq c_1}\ldiv_c(f_p)\Phi_c. \nonumber
\end{align}
An adjacent cube in the y-direction will have a term as $f_p^{-\tilde{y}}\Phi_{c_2}$.
Since they are the same plaquette, namely $f_p^{y}$, combining two terms from the adjacent cubes, we have,
\begin{align}
    f_p^{+y}\Phi_{c_1}-f_p^{-\tilde{y}}\Phi_{c_2}&=f_p^{y}\bigl(\Phi_{c_1}-\Phi_{c_2}\bigl) \nonumber \\
                                             &=-f_p^{y}\text{grad}_p(\Phi_c)
\end{align}
Note that $\lagr_{bE}^{\Phi}$ is a term concerning about potential $\Phi_c(t)$, which can be recasted into $U^{\Phi}_\m(c)=-g_{bE}\Phi_c Q_\m$.
In order to get the Newtonian force,
\begin{align}
    F_\m^{E}(t)&=-\grad_p U^{\Phi}_\m(c) \nonumber \\
          &=-\frac{1}{a}\text{grad}_p\bigl(U^{\Phi}_\m(c)\bigl) \nonumber \\
          &=-\frac{g_{bE}}{a}Q_\m E_p(t) \label{Monopole force due to E}
\end{align}
With this, we infer that the magnetic monopole has a physical electric charge of $-(g_{bE}/{a})Q_\m$.
We run a similar argument for the $\lagr^A_{bE}$ to verify our result of $q=-(g_{bE}/a)Q_\m$.
We start with the $\lagr^A_{bE}$ part as given in Eq.\eqref{L coupling of b and E},
\begin{align}
    \lagr^A_{bE}&=\sum_p g_{bE}\frac{dg_p}{dt}A_p-\sum_p g_{bE}\frac{d}{dt}\bigl(g_pA_p\bigl) \nonumber \\
                      &=-\sum_p g_{bE}(2\pi\frac{dM_p}{dt}+\lcurl_p e_l)A_p+\cancel{\text{total time derivative}} \label{L_bE}
\end{align}
Above, we used the lattice version of Eq.[\ref{Faraday's law}].  
Denoting $2\pi\dot{M}_p$ as $j_p^m$, we label this as the magnetic monopole current. Note, that it has the dimensions of $(v^m_p/a)Q_\m$. The action for a single monopole is calculated as
\begin{align}
    \mathcal{S}&=\int_{t_i}^{t_f} \lagr^A_{bE} dt \nonumber \\
               &=-\frac{g_{bE}}{a}Q_\m\int_{t_i}^{t_f} v_p^mA_p dt \nonumber \\
               &=-\frac{g_{bE}}{a}Q_\m\oint A_pdx        \label{Aharanov Bohm argument}
\end{align}
The above derivation confirms that we reach the same conclusion of the magnetic monopole having a physical charge of $-(g_{bE}/{a})Q_\m$ when we have an applied E field using an Aharanov Bohm argument,.\\
To obtain the force acting on the magnetic monopole due to the $g$ field, we inspect the $\lagr_{g_p}$ term of the Lagrangian. 
To see this, we treat a test monopole (i.e. one end of a sufficiently long Dirac string) in a background of fluctuations of other monopoles.
Splitting the smooth monopole field leads to,
\begin{align}
    \lagr_{g_p}=\sum_p\frac{(g_p^o)^2}{2\mu}+\frac{(g_p^m)^2}{2\mu}+ \frac{g_p^o g_p^m}{\mu}. \label{Lagrangian for g field}
\end{align}
The $(g_p^m)^2/2\mu$ term can be attributed to background fluctuations, while the $(g_p^o)^2/2\mu$ accounts for the self energy of the monopole.
The object of interest to us is the $(g_p^m.g_p^o)/\mu$ term which tells us about the interaction of monopole with the fluctuations.
% To obtain the force acting on the magnetic monopole due to the $g$ field, we look at the $g_p^mg_p^o/\mu$ term in Eq.[\ref{Lagrangian for g field}].
We define the test charge and background fluctuations along with their fields as follows,
\begin{align}
    \begin{aligned}
        g_p^m=-\text{grad}_p(\phi_c)
        \quad
        \ldiv_c(g_p^m)=Q_\m^{bg}(c)
        \quad
        \ldiv_c(g_p^o)=Q_\m^{\text{test}}(c)
    \end{aligned}
\end{align}
The term in $\lagr_{g_p}$ can be seen as,
\begin{align}
    U^{g_p}_\m&=\sum_p \frac{g^o_p(-\text{grad}_p(\phi_c))}{\mu} \nonumber \\
             &=\sum_c \frac{\ldiv_c(g_p^o)\phi_c}{\mu}\nonumber \\
             &=\sum_c \frac{Q_\m^{\text{test}}\phi_c(c)}{\mu} \nonumber
\end{align}
where we have used the fact that,$\sum_c \ldiv_c(f_p)\Phi_c=-\sum_p f_p\text{grad}_p(\Phi_c)$. The force is estimated as,
\begin{align}
    F^g_\m&=-\frac{1}{a}\text{grad}_p(U^{g_p}_m) \nonumber \\
                     &=\frac{Q_\m^{\text{test}}}{a}\frac{g_p^m}{\mu} \label{Monopole force due to g}
\end{align}
Combining the contribution of force from Eq.\eqref{Monopole force due to E} and \eqref{Monopole force due to g}, we obtain Eq.[\ref{Lorentz force equation}].
%[%%%%%%%%%%%%%%%%%%%%%%%%%%%%%%%%%%%%%%%%%%%%%%%%%%%%%%%%%%%%%%%%%%%%%%%%%%%%%%%%%%%%
\section{Comparison of speed and number density of the monopole and emergent photons}
\label{appendixA3}
In order to calculate the number densities of the monopoles, we will assume they are bosons having quadratic dispersion, i.e. $\epsilon_k^m=\hbar^2k^2/2m+\Delta_\m$ and two polarisations owning to $Q_\m=\pm2\pi$.
\begin{align}
    n_\m=2\int_{k\leq1/a}\frac{1}{e^{\beta\epsilon_k}-1} \frac{d^3k}{(2\pi)^3} \nonumber
\end{align}
In low-temperature limit i.e. $\beta\Delta_\m\gg1$, we have the following thermally activated number density,
\begin{align}
    n_\m=\frac{2}{\Lambda_T^3}e^{-\beta\Delta_\m}
\end{align}
Here $\Lambda_T^2=2\pi\beta\hbar^2/m$ denotes the thermal wavelength. For the emergent photons, we replace the $\epsilon_k^{ph}=\hbar ck$, since they are gapless and have linear dispersion.
\begin{align}
    n_{ph}=2\int_{k\leq1/a}\frac{1}{e^{\beta\hbar ck}-1}\frac{d^3k}{(2\pi)^3}\approx \frac{2}{\pi^2}\biggl(\frac{1}{\beta\hbar c}\biggl)^3
\end{align}
At low temperature, the $n_\m$ is exponentially suppressed by $\Delta_\m$ compared to $\sim T^3$ of $n_{ph}$. This justifies our assumption of treating the monopole as a test charge in the bath of the emergent photons.\\
In order to prove that the monopoles are slow compared to the photons, we take a look at the averaged kinetic energy,
\begin{align}
    \frac{\text{K.E.}}{V}=n_\m\frac{1}{2}m\langle v^2\rangle&=\int_{k\leq 1/a}2\frac{\hbar^2k^2/2m}{e^{\beta e_k}-1}\frac{d^3k}{(2\pi)^3} \nonumber\\ 
                         &\approx2 \frac{\hbar^2}{2m}6\pi\frac{e^{-\beta \Delta_\m}}{\Lambda_T^5}
\end{align}
To evaluate the average velocity per monopole, we have,
\begin{align}
    \sqrt{\frac{2 (\text{K.E/V)}}{m n_\m}}&=\sqrt{\frac{3\kB T}{m}} \nonumber\\
\end{align}
At low temperature, since the speed of the emergent photon is constant, they move faster than the monopoles,
\begin{align}
    \lim_{T\rightarrow 0}\frac{\sqrt{\langle v^2\rangle}}{c}=0
\end{align}
%%%%%%%%%%%%%%%%%%%%%%%%%%%%%%%%%%%%%%%%%%%%%%%%%%%%%%%%%%%%%%%%%%%%%%%%%%%%%%%%%%%%%
\section{Derivation of Electric susceptibility of a 3D $U(1)$ spin liquid slab}
\label{appendixA4}
\begin{figure}[t]
\includegraphics[width=0.25
\columnwidth,angle=0,clip=true]{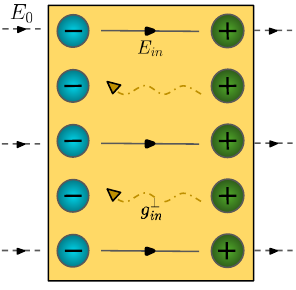}
 \caption{Slab of 3D $U(1)$ QSL material in an external electric field $E_0$, showcasing build up of a screening layer of monopoles at the boundaries. Different colors indicate monopoles of opposite charge.}
\label{Fig3}   
\end{figure}

In order to probe the dielectric nature of a 3D $U(1)$ quantum spin liquid candidate, we imagine placing a slab of the material between two conducting plates, effectively creating a capacitor. 
In other words, we place the system in a uniform external electric field $\ve{E}_0$, which in turn sets up an electric field $\ve{E}_\text{in}$ inside the material as shown in Fig.\ref{Fig3}. 
The dielectric properties of the material can be inferred from the relationship between $\ve{E}_\text{in}$ and $\ve{E}_0$. 

Following classical electrodynamics, by computing $\oint\ve{E}\cdot\ve{d}l$ along a vanishingly small loop constructed across the surface of the material, we find that the parallel component is continuous, $\ve{E}^{\Vert}_\text{in}=\ve{E}^{\Vert}_0$. 
Furthermore, considering the electric flux $\oint\ve{E}\cdot\ve{d}S$ through a Gaussian pill box across the surface, we establish that the normal component is discontinuous, i.e.,
    \begin{equation}\label{eq:E_from_rho}
        E^{\perp}_\text{out}-E^{\perp}_\text{in}=\frac{1}{\epsilon_0}\rho_{2D},
    \end{equation}
where $\rho_{2D}$ is the surface charge density of the material. 
In the geometry of Fig.\ref{Fig3}, $E^{\perp}_\text{out}=E_0$ and $\ve{E}^{\Vert}_0=0$. 

The emergent $g$ field inside the material also satisfies effective Maxwell equations, namely
    \begin{align}
        a\,\curl\left[\mu^{-1} g-\gbE\,E\right]&=
        \epsilon\,\dot{e}+\geB\,\dot{B}, \label{eq:one}
  \\
        a\,\div{g}&=Q_\m.\label{eq:two}
    \end{align}
We run the same argument as before for the physical electric fields on the emergent $g$ field, with the tacit understanding that the latter only exists inside the material, i.e. $g_\text{out}=0$. 
From Eq.~\eqref{eq:one} we find for the parallel component,
    \begin{equation}
        \mu^{-1}(g^{\Vert}_\text{out}-g^{\Vert}_\text{in})=\gbE(\ve{E}^\Vert_\text{out}-\ve{E}^\Vert_\text{in})
        \implies 
        g^{\Vert}_\text{in}=0,
    \end{equation}
i.e. the parallel component of $g$ vanishes.
Recalling that a monopole carries physical electric charge of $q=-Q_\m\gbE/a$, we can recast Eq.~\eqref{eq:two} as $\div{g}=-q/\gbE$. 
Integrating this equation over a vanishingly small Gaussian pillbox of the size of the unit cell (volume $a^3$), we get
    \begin{equation}
        \int\dd{V}\div{g}=-\frac{1}{\gbE}\int\dd{V}q
        \implies
        (g^\perp_\text{out}-g^{\perp}_\text{in})a^2=-\frac{a^3}{\gbE}\left(a^2\rho_{2D}\right),
    \end{equation}
which implies that the emergent field is connected to the surface charge density as,
    \begin{equation}\label{eq:g_from_rho}
        g^\perp_\text{in}=\frac{a^3}{\gbE}\rho_{2D}.
    \end{equation}
%%%
%%%
In equilibrium, the monopoles experience no net force inside the material, which from Eq.\eqref{Lorentz force equation} implies,
\begin{align}
    F_{m}=\frac{Q_\m}{a}\left(\frac{g_\text{in}^{\perp}}{\mu}-\gbE\,E_\text{in}^{\perp}\right)=0
    \implies
    g_\text{in}^{\perp}=\mu\,\gbE\,E_\text{in}^{\perp}.
    \label{eq:g_from_Ein}
\end{align}
Now using Eqs.~\eqref{eq:g_from_rho} and \eqref{eq:g_from_Ein} in Eq.~\eqref{eq:E_from_rho}, we establish the electric field inside the material is 
\begin{align}\label{Electric field inside dielectric}
    E_\text{in}=E_0\left[1+\frac{1}{\epsilon_0}\frac{(\gbE/a)^2}{\mu^{-1} a}\right]^{-1},
\end{align}
i.e. the applied electric field is partially screened inside the material. 
It is useful to compare this result to the screening behavior of usual insulators. 
For an insulator with electric susceptibility $\chi_E$, applied electric fields get screened as 
$E_\text{in}=E_0/(1+\chi_E)$. 
Comparing with Eq.\eqref{Electric field inside dielectric}, the electric susceptibility of the candidate material is given by $\chi_E=\epsilon_0^{-1}\mu\,\gbE^2/a^3$. 

Furthermore, if we drive a usual insulator with a spatially uniform oscillating electric field $E_0(t)=\Re[E_0(\omega)\mathrm{e}^{-\ii\omega t}]$, the polarization inside the material also oscillates $P(t)=\Re[P(\omega)\mathrm{e}^{-\ii\omega t}]$, where $P(\omega)=\epsilon_0\chi_E E_\text{in}(\omega)$ from the definition of susceptibility. 
The corresponding bound current is $J=\partial P/\partial t=-\ii\omega\epsilon_0\chi_E E_\text{in}(\omega)$. If we define the conductivity of an insulator as $J(\omega)=\sigma_\text{ins}E_\text{in}(\omega)$, we can read off $\sigma_\text{ins}(\omega)=-\ii\omega\epsilon_0\chi_E$. 
Thus in the low frequency limit ($\omega\ll \omega_p$) we expect the 3D $U(1)$ QSL candidate to have an insulating response in the conductivity given by $\sigma_\text{ins}(\omega)=-\ii\omega\,\mu\,\gbE^2/a^3$.  
This reaffirms our identification of the insulating part of resistivity in the Ioffe-Larkin rule in Eq.\eqref{Ioffe Larkin rule} of the main text.
 %%%%%%%%%%%%%%%%%%%%%%%%%%%%%%%%%%%%%%%%%%%%%%%%%%%%%%%%%%%%%%%%%%%%%%%%%%%%%%%%%%%%%
\section{Derivation of the Thomson scattering cross section}
\label{appendixA5}
This section presents a detailed derivation of the Thomson scattering cross section from the Maxwell's equations of the Quantum Spin ice.
The following Maxwell equations along with the force equation in absence of the external probe fields are;
\begin{align}
    a\curl{\ve{g}}&=\mu\epsilon\dot{\ve{e}} \label{Amp Maxwell's law wo ext}\\
    a\curl{\ve{e}}&=-[\dot{\ve{g}}+\ve{j_\m}] \label{Faraday's law wo ext}\\
    \epsilon a \div{\ve{e}}&=0 \label{Gauss law wo ext} \\
    a\div{\ve{g}}&=Q_\m \label{Magnetic gauss law wo ext} \\
    \ve{F}&=\frac{Q_\m}{a\mu}\ve{g} \label{Force Eqn wo ext}
\end{align}
Since $\div{\ve{e}}=0$, we can define $\ve{e}\equiv \curl{\ve{f}}$ by using Helmholtz theorem. Putting this in Eq.\eqref{Amp Maxwell's law wo ext}, we get,
\begin{align}
    \curl{\bigl(\ve{g}-\frac{\mu\epsilon}{a}\frac{\partial{{\ve{f}}}}{\partial t}\bigl)=0}\\
\end{align}
Defining the term inside the curl as $-\grad{U}$, the forms of $\ve{g}$ and $\ve{e}$ are,
\begin{align}\begin{aligned}
    \ve{g}=-\grad{U}+\frac{\mu\epsilon}{a}\frac{\partial\ve{f}}{\partial t},
    \quad
    \ve{e}=\curl{{\ve{f}}} \label{Eqns Maxwell}
\end{aligned}
\end{align} 
Note that a similar treatment is done in context of deriving the time dependent Maxwell equations with the difference of $\ve{e}=-\grad{V}-{\partial \ve{A}}/{\partial t}$ and $\ve{B}=\curl{\ve{A}}$.
The absence classical magnetic monopoles encoded in $\div{\ve{B}}=0$ and us being in the low-energy limit ($\div{\ve{e}}=0$) are mathematically equivalent. 
 Putting the expressions of $\ve{g}$ and $\ve{e}$ into Eq.\eqref{Amp Maxwell's law wo ext} and Eq.\eqref{Faraday's law wo ext} respectively we get,
\begin{align}
\begin{aligned}
    \nabla^2{U}-\frac{\partial}{\partial t}\div{\biggl(\frac{\mu\epsilon}{a}{\ve{f}}\biggl)}
    &=-\rho_\m a^2
\\
    a\biggl[\biggl(\nabla^2{\ve{f}}-\frac{\mu\epsilon}{a^2}\frac{\partial^2{\ve{f}}}{\partial t^2}\biggl)-\grad{\biggl(\div{{\ve{f}}}}-\frac{1}{a}\frac{\partial U}{\partial t}\biggl)\biggl]
    &=\ve{j_\m}
\label{Differential eqns}
\end{aligned}
\end{align}
Here we have used $Q_\m(c)=\rho_\m a^3$. In order to solve these two differential equations for potentials, we need to consider the gauge freedom associated with ${\ve{f}}$ and $U$.
 We define $\ve{f'}=\ve{f}+\tilde{{\boldsymbol{\alpha}}}$ and $U'=U+\tilde{\beta}$.
The curl free condition of $\ve{e}$ leads us to write $\tilde{\boldsymbol{\alpha}}=\grad{\lambda}$ for some potential $\lambda$.
Doing the same with $U'=U+\tilde{\beta}$ leads us to the condition $\div{\bigl(\tilde{\beta}-(\mu\epsilon)a^{-1}{\partial \lambda}/{\partial t}}\bigl)=0$.
The following are the allowed gauge transformations,
\begin{align}\begin{aligned}
    \ve{f'}&=\ve{f}+\grad{\lambda}\\
    U'&=U+\frac{\mu\epsilon}{a}\frac{\partial \lambda}{\partial t}
\end{aligned}
\end{align}
Note that we have a extra factor of $(\mu\epsilon){a}^{-1}$ in $U$ compared to the usual $V'=V-{\partial \lambda}/{\partial t}$ in electrodynamics.
To solve the above Eq.\eqref{Differential eqns} differential equations, we use the Lorentz-like gauge $\grad{.\ve{f}}={a^{-1}}{\partial U}/{\partial t}$ to simplify the problem.
The resulting PDE equations we get are,
\begin{align}
    \begin{aligned}
        \nabla^2{U}-\frac{\mu\epsilon}{a^2}\frac{\partial^2U}{\partial t^2}=\rho_\m a^2,
        \quad
        a\biggl[\nabla^2{\ve{f}}-\frac{\mu\epsilon}{a^2}\frac{\partial^2 \ve{f}}{\partial t^2}\biggl]=\ve{j_\m} \nonumber
    \end{aligned}
\end{align}
Defining $\nabla^2-{(\mu\epsilon)}{a^{-2}}\partial^2_{tt}\equiv \square^2$ as the d'Alembert operator, we write the PDE's as,
\begin{align}
    \begin{aligned}
        \square^2U=-\rho_\m a^2,
        \quad
        \square^2\ve{f}=\frac{\ve{j_\m}}{a}
    \end{aligned}
\end{align}
A look at the d'Alembert's operator tells us that the effective speed of emergent photons are $c=(\mu\epsilon)^{-1/2}a$.
Since we are dealing with a dynamic case, we have to treat these equations using Li\'{e}nard Wiechert potential assumptions, i.e by invoking \textit{retarded time} $t_r=t-{\mathcal{r}}/{c}$. 
The solution to satisfy these Poisson's equation can be given by,
\begin{align}
    \begin{aligned}
    U(r',t_r)=\frac{a^2}{4\pi}\int_\mathcal{V} \frac{\rho_\m(r',t_r)}{\mathcal{r}} d\tau',
    \quad
    \ve{f}(r',t_r)=\frac{1}{4\pi a}\int_\mathcal{V}\frac{\ve{j}_\m(r',t_r)}{\mathcal{r}}d\tau'\nonumber
    \end{aligned}
\end{align}
Taking into account the retardation factor, the potentials for a single moving charge $Q_\m$ becomes,
\begin{align}
    \begin{aligned}
        U(r',t_r)=\frac{a^2}{4\pi}\frac{Q_\m}{1-\mathcal{\hat{r}}.\ve{v}/c},
        \quad
        \ve{f}(r',t_r)=-\frac{a}{4\pi}\frac{Q_\m\ve{v}}{1-\mathcal{\hat{r}}.\ve{v}/c}
    \end{aligned}
\end{align}
Putting the forms back into Eq.\eqref{Eqns Maxwell}, we recover the $\ve{g}$ and $\ve{e}$ as,
\begin{align}
    \begin{aligned}
        \ve{g}(r,t)
        &=\frac{a^2}{4\pi}\frac{Q_\m \mathcal{r}}{(\boldsymbol{\mathcal{r}}.\ve{u})^3}\bigl[(c^2-{v}^2)\ve{u}+\boldsymbol{{\mathcal{r}}}\cross(\ve{u}\cross \ve{a})\bigl]
    \\
        \ve{e}(r,t)
        &=-\frac{a}{4\pi}\frac{Q_\m c}{(\boldsymbol{\mathcal{r}}.\ve{u})^3}\bigl(\boldsymbol{\mathcal{r}}\cross\bigl[(c^2-v^2)\ve{u}+\boldsymbol{\mathcal{r}}\cross(\ve{u}\cross \ve{a})\bigl]\bigl)
    \end{aligned}
\end{align}
where we use $\ve{u}\equiv c\mathcal{\hat{r}}-\ve{v}$ and $\ve{a}$ as acceleration of the charged particle.
 More compactly, we can write $\ve{e(r,t)}=-{c}/{a}\bigl(\hat{\mathcal{r}}\cross\ve{g}(r,t)\bigl)$. 
Note the close resemblance $\ve{g}$ and $\ve{e}$ to that of $\ve{e}$ and $\ve{B}$ field of a point charge in Maxwell's electrodynamics, $B(\ve{r},t)=c^{-1}(\mathcal{\hat{r}} \cross E(\ve{r},t))$.
The first term in $\ve{g}$ i.e. $\bigl[(c^2-v^2)\ve{u}\bigl]$ is called the \textbf{velocity field}.
This term falls off as $\mathcal{r}^{-2}$ and does not contribute to Thomson scattering.
In static cases, this reduces to the generalized Coulomb field.
The second term (acceleration field), $\boldsymbol{\mathcal{r}}\cross (\ve{u}\cross\ve{a})$ falls off as $\mathcal{r}^{-1}$ and is the dominant term in electromagnetic radiation pressure.\\
In order to calculate the power radiated by a single charge we need to derive the Poynting vector.
Writing down the rate of work done on all charges in volume $\mathcal{V}$ in  terms of charge ($Q_\m=\int_{\mathcal{V}}\rho_\m d\tau'$) and current densities (${\ve{j_\m}}=a^2\rho_\m\ve{v}$), we have,
\begin{align}
    \frac{dW}{dt}=\frac{Q_\m.\ve{g}}{a\mu}.\ve{v}
                 =\int_{\mathcal{V}}\frac{\rho_\m \ve{g}}{a\mu}.\ve{v}d\tau'
                 =\int_{\mathcal{V}}\frac{\ve{g}.\ve{j_\m}}{a^3\mu}d\tau'
\end{align}
We use Eq.\eqref{Faraday's law wo ext}  to rewrite ${j_\m}$ and the vector identity $\div{(\ve{g}\cross\ve{e})}=e.(\curl{\ve{g}})-g.(\curl{\ve{e}})$
\begin{align}
    \ve{g}.\ve{j_\m}&=\ve{g}.(-\dot{\ve{g}}-a\curl{\ve{e}})\nonumber\\
                  &=-\ve{g}.\dot{\ve{g}}-a\ve{e}.\curl{\ve{g}}+a\div{(\ve{g}\cross\ve{e})}\nonumber\\
                  &=-\ve{g}.\dot{\ve{g}}-\mu\epsilon\ve{e}\dot{\ve{e}}+a\div{\ve{g}\cross\ve{e}}\nonumber\\
                  &=-\frac{1}{2}\partial_tg^2-\frac{1}{2}\mu\epsilon\partial_te^2+a\div{(\ve{g}\cross\ve{e})}
\end{align}
The first two terms can be seen as the energy stored in the $\ve{g}$ and $\ve{e}$ field. Defining the energy density stored in these fields as $u_\m=\bigl({\mu^{-1}}g^2+\epsilon e^2\bigl)/a^3$, we have,
\begin{align}
    \frac{1}{a^3}\frac{dW}{dt}=-\frac{d u_\m}{dt}-\oint_\mathcal{S}\frac{1}{a^2\mu}(\ve{e}\cross\ve{g}).da
\end{align}
This is the Poynting's theorem for conservation of energy for our version of Maxwell's equations. The second term represents the rate at which energy is dissipating through the surface $\mathcal{S}$. The energy flux or Poynting vector is defined as,
\begin{equation}
    \tilde{S}=\frac{1}{a^2\mu}(\ve{e}\cross\ve{g})
\end{equation}
In order to calculate the power radiated by the charge $P(r,t)=\oint_{\mathcal{S}}\tilde{S}.d\ve{a}$, we consider the only the acceleration field of the g field $\thicksim (a^2/4\pi)(Q_\m\mathcal{r})((\boldsymbol{\mathcal{r}}.\ve{u})^{-3}\bigl(\boldsymbol{\mathcal{r}}\cross(\ve{u}\cross\ve{a})\bigl)$
\begin{align}
    \tilde{S}=\frac{1}{a^2\mu}\bigl[(-\mathcal{\hat{r}}\cross\ve{g})\cross\ve{g}\bigl].\frac{c}{a}
             =\frac{c}{a^3\mu}\bigl[g^2\mathcal{\hat{r}}-(\mathcal{\hat{r}}.\ve{g}).\ve{g}\bigl] \nonumber  
\end{align}
Assuming the charge is instantaneously at rest, i.e. $\ve{u}=c\mathcal{\hat{r}}$, the contribution from the $(\mathcal{\hat{r}}.\ve{g})$ is 0 (since $\ve{g}\thicksim \bigl[\mathcal{r\cross}(\ve{u}\cross\ve{a})\bigl]$ is perpendicular to $\mathcal{\hat{r}}$).
Simplifying $\ve{g}$ for the radiation and putting it's value in $\tilde{S}$ we conclude,
\begin{align}
\begin{aligned}
\ve{g}_{rad}(r,t)=\frac{a^2Q_\m}{4\pi c^2}\frac{1}{\mathcal{r}}\bigl[(\mathcal{\hat{r}}.\ve{a})\mathcal{\hat{r}}-\ve{a}\bigl],
\quad
    \tilde{S}=\frac{c}{a^3\mu}\biggl(\frac{a^2Q_\m}{4\pi c^2}\frac{1}{\mathcal{r}^2}\biggl)^2\ve{a}^2\sin^2{\Theta}.\mathcal{\hat{r}} \nonumber
\end{aligned}
\end{align}
Here $\Theta$ denotes the angle between the acceleration of the charges particle $\ve{a}$ and $\mathcal{\hat{r}}$ as the observer.
 Integrating out the radial dependence, we are left with the angular dependence, giving us power radiated per solid angle,
\begin{align}
    \frac{dP}{d\Omega}=\frac{c}{a^3\mu}\biggl(\frac{a^2Q_\m}{4\pi c^2}\biggl)^2\ve{a}^2\sin^2\Theta
\end{align}
Using the force equation \eqref{Force Eqn wo ext}, we replace the acceleration of particle as $\ve{a}(t)=(Q_\m/a\mu m)\ve{g}(t)$. The scattering cross section can be written as,
\begin{align}
    \frac{d\sigma_s}{d\Omega}=\frac{dP/d\Omega}{\langle\tilde{S}\rangle_t}
\end{align}
The $\langle\tilde{S}\rangle_t$ denotes the time averaged Poynting vector, which turns out to be $(g_0^2/2)(c/(a^3\mu))$ assuming the charge moves negligible distance in one cycle of oscillation compared to the wavelength.
Plugging these values we get,
\begin{align}
    \frac{d\sigma_s}{d\Omega}=\frac{1}{(4\pi)^2}\biggl(\frac{Q_\m^2}{mc^2}\biggl)^2\biggl(\frac{a}{\mu}\biggl)^2\sin{\Theta} \nonumber
\end{align}
The factor of $\sin{\Theta}$ is purely a geometric factor due to polarization of incoming photon and the observer.
 We assume a photon of propagation vector $\hat{k}$ (taken along $\hat{z}$ here) and polarization $\hat{\epsilon}$ is scattered off the monopole which is momentarily at rest along $\mathcal{\hat{r}}$. The oscillating emergent field $\ve{g}(r,t)=\Re[g_0e^{\ii(\mathbf{k\cdot x}-\omega t)}]$ accelerates the monopole as $\ve{a}(t)$ which in turn leads to electromagnetic dipole radiation.
The photon radiated by this is dipole is the scattering phenomena known as the \textbf{Thomson Scattering}.
Averaging over all possible incident polarizations and denoting the angle between the incident and outgoing photon as $\psi$, the final Thomson cross-section we get is,
\begin{align}
    \frac{d\sigma_s}{d\Omega}=
    \left(\frac{Q_\m^2\mu^{-1}a}{4\pi mc^2}\right)^2\frac{1}{2}(1+\cos^2\psi)
\end{align}
%%%%%%%%%%%%%%%%%%%%%%%%%%%%%%%%%%%%%%%%%%%%%%%%%%%%%%%%%%%%%%%%%%%%%%%%%%%%%%%%%%%%%
\section{Boltzmann equation for magnetic monopole}
\label{appendixA6}
In order to derive the Boltzmann equation for the magnetic monopole in the bath of emergent photons, we treat the monopoles in the dilute limit.
Denoting the distribution of the monopoles in the phase space as $f_p(r,t)$, we have
\begin{align}
    \frac{\partial f_p}{\partial t}+\vec{v}.\nabla f+\frac{\vec{F}}{m}\nabla_pf_p=\frac{\partial f_p}{\partial t}\bigg\rvert_{coll}
\end{align}
We assume that they are randomly distributed through the bath (i.e. we have no spatial variations of $f_p(r,t)$) and are close to equilibrium, i.e. $f_p=f^{(0)}_p+\delta f_p$.
Since we are in the dilute limit, we will be using $1+f_p\approx f_p$ as an approximation.
Subsequent to these assumptions, the only surviving term is $\partial f_p/\partial t\big\rvert_{coll}$, which we express as,
\begin{align}
    \frac{\partial f}{\partial t}\bigg\rvert_{coll}&=\sum_{p'}W_{p'\rightarrow p}f_{p'}(1+f_p)-W_{p\rightarrow p'}f_p(1+f_{p'}) \nonumber \\
                                                   &=\sum_{p'}W_{p'\rightarrow p}f_{p'}-W_{p\rightarrow p'}f_p
\end{align}
Here we use the standard notation of $W_{p'\rightarrow p}$ denoting the scattering probability of the monopole from momentum $p'$ to $p$.
The factor of $1+f_p$ is the Bose enhancement factor.
Adding to this, we use the test-particle approximation $\delta f_p=\delta_{p,p_0}$ which leads to the collision term as,
\begin{align}
    \frac{\partial f_p}{\partial t}= W_{p_0\rightarrow p}-\sum_{p'}W_{p\rightarrow p'}\delta_{p,p_0}
\end{align}
Since we are interested in treating this setup as a Drude model, we need to calculate the momentum relaxation $\big\langle\Delta v/\Delta t\big\rangle$.
Using the above arguments we get,
\begin{align}
    \frac{1}{m}\sum_p\frac{\partial f_p}{\partial t}p&=\sum_p pW_{p_0\rightarrow p}-\sum_{p'}\biggl(\sum_p\delta_{p,p_0}W_{p\rightarrow p'}\biggl) \nonumber \\
                     &=\sum_p pW_{p_0\rightarrow p}-\sum_{p'}p_0W_{p_0\rightarrow p'}\nonumber \\
                     &=\sum_p(p-p_0)W_{p_0\rightarrow p}
\end{align}
In the last line, we relabeled the dummy variable $p'$ to $p$.
The $W_{p_0\rightarrow p}$ factor can be calculated using Fermi's Golden Rule i.e. ,
\begin{align}
    W_{p_0\rightarrow p}\sim \frac{2\pi}{\hbar}|\langle f|V_{int}|i\rangle|^2\delta(\epsilon_i-\epsilon_f)
\end{align}
Here, $\langle f|V_{int}|i\rangle$ denotes the scattering matrix with $|i\rangle=|N_q,N_{q'}\rangle$ as the initial state, $|f\rangle=|N_q-1,N_{q'}+1\rangle$ the final state and $V_{int}$ as the interaction potential between the bath and the monopole.% and $\sigma(\Psi)$ is the Thomson scattering cross-section we calculated in the previous section. 
We model this potential as $V\sim \hat{a}^\dag_{q'}\hat{a}_q$ and the states as $|\{N_q\}\rangle=\Pi_q(N_q!)^{-1/2}(\hat{a}^\dag_q)^{N_q}|0\rangle$.
Effectively, the interaction of the monopole with the bath is  annihilating a photon of momentum of $q$ from $|N_q\rangle$ and creating a photon with momentum $q'$ in the final state of $|N_{q'}\rangle$.
The occupation factors are accompanied with the Thomson differential cross-section $\frac{d\sigma_s}{d\Omega}$ that we calculated in the previous section. 
The matrix element is then,
\begin{align}
    |\langle f|V_{int}|i\rangle|^2=
    |\langle f|\hat{a}^\dag_{q'}\hat{a}_q|i\rangle|^2 \frac{d\sigma_s}{d\Omega}
    % \nonumber\\
      =|\langle N_{q'}+1,N_q-1|\hat{a}^\dag_{q'}\hat{a}_q|N_{q'},N_q\rangle|^2 
      \frac{d\sigma_s}{d\Omega}
    % \nonumber\\
      =(N_{q'}+1)(N_q)\frac{d\sigma_s}{d\Omega}
\end{align}
We model the $N_q$ photons as blackbody radiation intensity, $I_\nu$.
In order to simulate this, we make a transformation of the photon of frequency $\nu$ and moving in the direction $\ve{n}$ in the laboratory frame, to the rest frame of the monopole.
By this assumption, the isotropic radiation field of specific intensity moving with velocity $\ve{v}$ becomes,
\begin{align}
    \begin{aligned}
    I_0(\nu_0,\ve{n}_0)=\frac{2h\nu_0^3}{c^3}\frac{1}{e^{h\nu_0/\kB T}-1}\biggl(1-\frac{1}{c}\frac{h\nu_0/\kB T}{1-e^{-h\nu_0/\kB T}}\ve{n}_0.\ve{v}\biggl), \quad\quad
    \nu_0 \rightarrow \nu(1-\frac{1}{c}\ve{n}.\ve{v})
    \end{aligned}
\end{align}
We have denoted the angle between the photon and the monopole as $\cos(\mathcal{v})$.
The Thomson scattering process $(\nu_0,\ve{n}_0)\rightarrow(\nu_0,\ve{n}_0')$ leads to change in velocity in the direction of $\ve{v}$ as, 
\begin{align}
    \frac{p-p_0}{m}=\frac{h\nu_0}{mc}\bigl(\cos\theta_0-\cos\theta_0'\bigl) \nonumber 
\end{align}
Here $\theta_0$ and $\theta_0'$ denote the angle between the velocity $\ve{v}$ of the monopole and the incoming and outgoing photon $\ve{n}_0$ and $\ve{n}_0'$ respectively.
Incorporating the $N_q$ and $N_{q'}$ as incoming and outgoing flux density of the photons, the velocity change per unit time due to Thomson Scattering is recovered as given in Ref.\cite{Oxenius1986},
\begin{align}
    \bigg\langle\frac{\Delta v}{\Delta t}\bigg\rangle=\int_{\nu_0,\Omega_0,\Omega_0'}I_0(\nu_0,\ve{v}_0)\biggl[1+\frac{c^2}{2h\nu_0^3}I_0(\nu_0,\ve{v}_0')\biggl]\frac{1}{h\nu_0}\frac{d\sigma_s}{d\Omega}(\psi)
    \frac{h\nu_0}{mc}\bigl(\cos\theta_0-\cos\theta_0'\bigl)
    d\nu_0d\Omega_0d\Omega_0'
\end{align}
The angle $\psi$ in $d\sigma_s/d\Omega$ denotes the angle between the incoming and outgoing photon, and is given by $\cos\psi=\ve{n}_0.\ve{n}_0'$ where $\ve{n}_0$ and $\ve{n}_0'$ are unit vectors,
\begin{align}
    \ve{n}_0=(\sin\theta_0\cos\phi_0,\sin\theta_0\sin\phi_0,\cos\theta_0) 
    \quad
    \ve{n}_0'=(\sin\theta_0'\cos\phi_0',\sin\theta_0'\sin\phi_0',\cos\theta_0')
\end{align}
We neglect $\mathcal{O}(\ve{v}^2/c^2)$ after integrating over $d\Omega_0$ and $d\Omega_0'$.
After inserting the expression of Planck distribution from the $I_o(\nu_0)$ we write the remaining integral of $\nu_0$ in terms a new variable $\zeta=h\nu_0/\kB T$ as,
\begin{align}
    \bigg\langle\frac{\Delta v}{\Delta t}\bigg\rangle=-v\frac{Q_\m^4}{m^3c^8}\biggl(\frac{a}{\mu}\biggl)^2\frac{2^3}{3^2}\frac{(\kB T)^4}{h^3}\int_0^{\frac{W_\phi}{\kB T}}\frac{\zeta^4e^\zeta}{(e^\zeta-1)^2}d\zeta,
\end{align}
where we have defined $W_\phi\coloneq \hbar c/a$ as the bandwidth of the photons. 
From the definition $\langle{\Delta v}/{\Delta t}\rangle=-\Gamma v$, the dissipation rate is then
    \begin{equation}
        \Gamma(T)=
        \frac{2^4\pi}{3^2}
        \left(\frac{Q_\m}{2\pi}\right)^4
        \left(\frac{\mu^{-1}}{mc^2}\right)^2
        \frac{\hbar}{ma^2}
        \left(\frac{\kB T}{W_\phi}\right)^4
        \int_0^{\frac{W_\phi}{\kB T}}\frac{\zeta^4e^\zeta}{(e^\zeta-1)^2}d\zeta.
    \end{equation}
% The last integral is calculated using a identity involving Riemann zeta function,
% \begin{align}
%     \int_0^{\infty}\frac{x^se^x}{(e^x-1)^2}dx=\Gamma(s+1)\Gamma(s), \quad s>1 \nonumber
% \end{align}
At low temperatures $\kB T\ll W_\phi$, the upper limit in the last integral can be set to $\infty$, and the resulting integral is exactly evaluated to be $4\pi^4/15$, which leads to the result in Eq.~\eqref{Relaxation rate} of the main text,
\begin{align}
    \Gamma\approx
    %\frac{2^5\pi^4}{3^3.5}\frac{Q_\m^4}{m^3c^8}\biggl(\frac{a}{\mu}\biggl)^2\frac{(\kB T)^4}{h^3}
    \frac{4\pi\,Q_\m^4}{135}
        \left(\frac{\mu^{-1}}{mc^2}\right)^2
        \frac{\hbar}{ma^2}
        \left(\frac{\kB T}{W_\phi}\right)^4
\end{align}
%%%%%%%%%%%%%%%%%%%%%%%%%%%%%%%%%%%%%%%%%%%%%%%%%%%%%%%%%%%%%%%%%%%%%%%%%%%%%%%%%%%%%
\section{Physical Electric charge of monopoles in Dipolar-Octupolar Quantum Spin ice}
\label{appendixA7}
\label{sec:do_monopole_charge}

This section discusses which dipolar--octupolar (DO) quantum spin-ice phases in ideal pyrochlore compounds have emergent magnetic monopoles with physical electric charge.  Answering this question boils down to determining whether the emergent magnetic field, \(\emg\), in these states has the same transformation rules under crystalline symmetries and time reversal as the physical electric field \(\physE\).

\subsection{Summary of symmetry transformations in DO-QSI}
\label{subsec:symmetry_input_DO_revised}

Our analysis here builds upon the seminal work of Ref.~\cite{Huang.Hermele_PRL14_QuantumSpinIcesb}, which derived the transformation rules for the effective pseudo-spin \(1/2\) operators defined in the local tetrahedral coordinate frame.  In this frame, \(\tau^x\) and \(\tau^z\) transform like the local \(z\)-component of a magnetic dipole, while \(\tau^y\) transforms like a magnetic octupole component.  As we will show, the symmetry that creates a crucial distinction between the Dipolar (D) and Octupolar (O) $U(1)$ quantum spin liquids is the mirror \(\Msym\), under which
\begin{equation}
    \Msym:\quad
    \tau^{x,z}_r \mapsto -\tau^{x,z}_{\Msym r},
    \qquad
    \tau^y_r \mapsto +\tau^y_{\Msym r} .
    \label{eq:tau_mirror_DO_revised}
\end{equation}
As shown in Ref.~\cite{Huang.Hermele_PRL14_QuantumSpinIcesb}, for a pyrochlore compound of the form \(A_2B_2O_7\), where \(A\) is a trivalent rare earth with a partially filled \(4f\) shell and \(B\) is a non-magnetic transition metal (such as \(\mathrm{Ce}_2\mathrm{Zr}_2\mathrm{O}_7\)), the symmetry-allowed nearest-neighbor coupling Hamiltonian has the form
\begin{equation}
    H_{\rm XYZ}=\sum_{\langle rr'\rangle}
    \tilde J_x\tilde\tau^x_r\tilde\tau^x_{r'}
    +\tilde J_y\tilde\tau^y_r\tilde\tau^y_{r'}
    +\tilde J_z\tilde\tau^z_r\tilde\tau^z_{r'} .
    \label{eq:xyz_DO_revised}
\end{equation}
Since \(\tilde\tau^x\) and \(\tilde\tau^z\) are linear combinations of \(\tau^x\) and \(\tau^z\), they have the same symmetry transformations, while \(\tilde\tau^y=\tau^y\).  It will be Dipolar (D) if either \(\tilde J_x\) or \(\tilde J_z\) is the strongest easy-axis anisotropy, and it is Octupolar (O) if \(\tilde J_y\) is the strongest.

In these easy-axis limits, the model maps to a compact $U(1)$ gauge theory on the dual diamond lattice, where the emergent lattice electric fields and vector potentials, denoted by \(e_{\rr\rrp}\) and \(a_{\rr\rrp}\), can be identified with lattice pseudo-spin operators as follows:
\begin{subequations}
\begin{align}
    D_z:\quad & \tilde\tau^z_r=e_{\rr\rrp},
    & \tilde\tau^+_r&=\tilde\tau^x_r+i\tilde\tau^y_r\simeq e^{i a_{\rr\rrp}},
    \label{eq:gauge_map_Dz_DO_revised}\\
    D_x:\quad & \tilde\tau^x_r=e_{\rr\rrp},
    & \tilde\tau^+_r&=\tilde\tau^y_r+i\tilde\tau^z_r\simeq e^{i a_{\rr\rrp}},
    \label{eq:gauge_map_Dx_DO_revised}\\
    O:\quad & \tilde\tau^y_r=e_{\rr\rrp},
    & \tilde\tau^+_r&=\tilde\tau^z_r+i\tilde\tau^x_r\simeq e^{i a_{\rr\rrp}}.
    \label{eq:gauge_map_O_DO_revised}
\end{align}
\end{subequations}
The \(D_z\) and \(D_x\) phases are equivalent for the present symmetry question, so below they are collectively denoted by D QSI.

\begin{table}[h]
\caption{Transformation rules relevant for linear couplings between emergent and physical fields.}
\label{tab:transformation_summary_DO_revised}
\begin{center}
\renewcommand{\arraystretch}{1.18}
\begin{tabular}{lccc}
\toprule
Field & Inversion \(\Id\) & Mirror \(\Msym\) & Time reversal \(\cT\) \\
\midrule
\(\eme_D\) & pseudo-vector & pseudo-vector & odd \\
\(\emg_D\) & vector & vector & even \\
\(\eme_O\) & pseudo-vector & vector & odd \\
\(\emg_O\) & vector & pseudo-vector & even \\
physical \(\physE\) & vector & vector & even \\
physical \(\physB\) & pseudo-vector & pseudo-vector & odd \\
\bottomrule
\end{tabular}
\end{center}
\end{table}

Table~\ref{tab:transformation_summary_DO_revised} summarizes our main conclusions, which we detail in the next subsections, for the transformation laws of the emergent electric and magnetic fields in the dipolar and octupolar phases, as well as the physical electric and magnetic fields.  The vectors (a.k.a. ``polar-vectors'') transform in the same way as a space point, while the pseudo-vectors (a.k.a. ``axial-vectors'') pick an extra minus sign.

We now give two complementary derivations of Table~\ref{tab:transformation_summary_DO_revised}.  The first is a continuum gauge-field argument based on the transformation of \(\eme\), \(\ema\), and \(\emg\).  The second is a microscopic lattice argument based directly on the six-spin plaquette operator that measures the compact gauge flux.

\subsection{Continuum argument from \(\eme\) to \(\ema\) to \(\emg\)}
\label{subsec:continuum_argument_DO_revised}

The microscopic electric field \(e_{\rr\rrp}\) is the pseudo-spin component entering the spin-ice rule.  Its transformation under proper rotations is the one given in Ref.~\cite{Huang.Hermele_PRL14_QuantumSpinIcesb}.  Time reversal is not differentiating: all pseudo-spin components are time-reversal odd, hence both \(\eme_D\) and \(\eme_O\) are time-reversal odd.  Inversion acts trivially on the local pseudo-spin in Ref.~\cite{Huang.Hermele_PRL14_QuantumSpinIcesb} and gives the emergent electric field the inversion parity of a pseudo-vector in both D and O cases.  These rules are summarized in the first and last column of Table~\ref{tab:transformation_summary_DO_revised}.

The mirror $\Msym$ is the crucial operation.  Let $\Mmat$ denote the usual $3\times3$ mirror matrix acting on ordinary vectors.  For an ordinary vector, the mirror action is \(\mathbf v\mapsto \Mmat\mathbf v\); for a pseudo-vector, it is \(\mathbf v\mapsto -\Mmat\mathbf v\).  From Eq.~\eqref{eq:tau_mirror_DO_revised}, the dipolar easy-axis pseudo-spin field is odd under \(\Msym\), while the octupolar easy-axis pseudo-spin field is even.  Omitting the transformed spatial argument, this gives
\begin{subequations}
\begin{align}
    \Msym:\quad &\eme_D\mapsto -\Mmat\eme_D,
    \label{eq:eD_mirror_DO_revised}\\
    \Msym:\quad &\eme_O\mapsto +\Mmat\eme_O.
    \label{eq:eO_mirror_DO_revised}
\end{align}
\end{subequations}
Thus \(\eme_D\) transforms like the physical magnetic field \(\physB\), whereas \(\eme_O\) has the mixed character listed in Table~\ref{tab:transformation_summary_DO_revised}: pseudo-vector under inversion, but ordinary vector under \(\Msym\).

The transformation of the vector potential is fixed by the continuum relation \(\eme=-\nabla\varphi-\partial_t\ema\).  Therefore \(\ema\) has the same spatial transformation character as \(\eme\), while its time-reversal parity is opposite: \(\ema\) is time-reversal even.  Consequently
\begin{subequations}
\begin{align}
    \Msym:\quad &\ema_D\mapsto -\Mmat\ema_D,
    \label{eq:aD_mirror_DO_revised}\\
    \Msym:\quad &\ema_O\mapsto +\Mmat\ema_O,
    \label{eq:aO_mirror_DO_revised}\\
    \cT:\quad &\ema_D,\ema_O\mapsto +\ema_D,+\ema_O .
    \label{eq:aT_DO_revised}
\end{align}
\end{subequations}

Finally, the smooth emergent magnetic field is the curl of the vector potential,
\begin{equation}
    \emg=\nabla\times\ema .
    \label{eq:g_curl_a_DO_revised}
\end{equation}
Taking a curl flips the ordinary-vector/pseudo-vector character under any improper spatial operation.  Hence in the D case, where \(\ema_D\) is a pseudo-vector under both inversion and \(\Msym\), the field \(\emg_D\) is an ordinary vector under both.  In particular
\begin{equation}
    \Msym:\quad \emg_D\mapsto +\Mmat\emg_D .
    \label{eq:gD_mirror_DO_revised}
\end{equation}
This is precisely the mirror transformation of the physical electric field.  Therefore
\begin{equation}
    \Msym:\quad
    \emg_D\cdot\physE \mapsto
    (\Mmat\emg_D)\cdot(\Mmat\physE)
    =\emg_D\cdot\physE,
    \label{eq:gD_E_allowed_DO_revised}
\end{equation}
so \(\emg_D\cdot\physE\) is not excluded by the mirror symmetry.

In the O case, \(\ema_O\) is a pseudo-vector under inversion but an ordinary vector under \(\Msym\).  Its curl is therefore an ordinary vector under inversion but a pseudo-vector under \(\Msym\):
\begin{equation}
    \Msym:\quad \emg_O\mapsto -\Mmat\emg_O .
    \label{eq:gO_mirror_DO_revised}
\end{equation}
This justifies the corresponding entries in Table~\ref{tab:transformation_summary_DO_revised}.  It also immediately forbids the monopole-charge coupling, because
\begin{equation}
    \Msym:\quad
    \emg_O\cdot\physE \mapsto
    (-\Mmat\emg_O)\cdot(\Mmat\physE)
    =-\emg_O\cdot\physE .
    \label{eq:gO_E_forbidden_DO_revised}
\end{equation}
Equivalently, \(\nabla\cdot\emg_D\) transforms as an ordinary scalar and may couple to the physical scalar potential \(\Phi\), while \(\nabla\cdot\emg_O\) has the wrong mirror parity and cannot couple linearly to \(\Phi\).  This is the continuum reason that the D monopole can carry physical electric charge whereas the O monopole is neutral in the ideal symmetry setting.  Notice also that Table~\ref{tab:transformation_summary_DO_revised} implies that neither in the D nor in the O case does the spin-ice-rule-violating charge, associated with the divergence of \(\eme\), carry net physical charge.

\subsection{Microscopic plaquette-flux argument}
\label{subsec:microscopic_loop_DO_revised}

The same conclusion can be achieved directly at the level of the lattice flux operator.  In the perturbative QSI regime of the XYZ model, the leading ring exchange around an elementary hexagonal plaquette \(\Gamma\) contains the alternating product
\begin{equation}
    O_\Gamma
    =\tilde\tau^+_1\tilde\tau^-_2\tilde\tau^+_3\tilde\tau^-_4\tilde\tau^+_5\tilde\tau^-_6
    \simeq \exp\left(i\sum_{\ell\in\Gamma} a_\ell\right)
    =e^{i b_\Gamma}.
    \label{eq:ring_operator_DO_revised}
\end{equation}
Here \(1,\ldots,6\) label the six pyrochlore sites on the hexagonal plaquette, or equivalently the six diamond links whose midpoints are those sites.  The loop made of tetrahedron-center links on the diamond lattice is generally slightly puckered in the three-dimensional embedding, whereas the six pyrochlore midpoints form the familiar planar hexagon.  This geometric distinction is useful when drawing the mirror operation, but it does not affect the transformation of the link operators in Eq.~\eqref{eq:ring_operator_DO_revised}.

We focus on the same mirror $\Msym$.  The local transformation rules of Eq.~\eqref{eq:tau_mirror_DO_revised} give, for the two dipolar cases,
\begin{subequations}
\begin{align}
    D_z:\quad
    \tilde\tau^+=\tilde\tau^x+i\tilde\tau^y
    &\mapsto -\tilde\tau^- ,
    \label{eq:Dz_tau_plus_DO_revised}\\
    D_x:\quad
    \tilde\tau^+=\tilde\tau^y+i\tilde\tau^z
    &\mapsto +\tilde\tau^- .
    \label{eq:Dx_tau_plus_DO_revised}
\end{align}
\end{subequations}
The extra minus sign in the \(D_z\) case appears on every link of the six-link product and therefore cancels.  After also accounting for the fact that a mirror reverses the oriented path used to define the gauge-basis product, the dipolar plaquette operator maps to the plaquette operator on the mirror-related hexagonal plaquette rather than to its Hermitian conjugate:
\begin{equation}
    D:\quad O_\Gamma\mapsto O_{\Msym\Gamma} .
    \label{eq:D_loop_transform_DO_revised}
\end{equation}
Hence the smooth flux component transforms as
\begin{equation}
    D:\quad g_\Gamma\mapsto +g_{\Msym\Gamma} .
    \label{eq:D_flux_even_DO_revised}
\end{equation}
The above is the same transformation law of the physical electric field under this mirror.

For the octupolar case,
\begin{equation}
    O:\quad
    \tilde\tau^+=\tilde\tau^z+i\tilde\tau^x
    \mapsto -\tilde\tau^+ .
    \label{eq:O_tau_plus_DO_revised}
\end{equation}
The overall signs again cancel in a six-link loop, but now the mirror action on the oriented loop gives the Hermitian-conjugate plaquette operator:
\begin{equation}
    O:\quad O_\Gamma\mapsto O^{\dagger}_{\Msym\Gamma} .
    \label{eq:O_loop_transform_DO_revised}
\end{equation}
Since \(O_\Gamma\simeq e^{i b_\Gamma}\), the smooth flux changes sign,
\begin{equation}
    O:\quad g_\Gamma\mapsto -g_{\Msym\Gamma} .
    \label{eq:O_flux_odd_DO_revised}
\end{equation}
Thus the octupolar lattice flux has the mirror character of a pseudo-vector, unlike the physical electric field.  Similar arguments to the one above can be done to verify the other transformations directly in the lattice model.

\section{Quality factors: resonant vs. onset regimes}\label{ap:regimes}

In this section, we examine the expression for conductivity in Eq.~\eqref{sigmaomega} of the main text. 
As before, we consider a charged fluid of $Q_\m=\pm2\pi$ monopoles at temperatures below the monopole gap $\Delta_\m$, with a thermally activated density $n_m$,
    \begin{equation}
        a^3n_\m(T)\approx 2\left(\frac{\kB T}{2\pi W_\m}\right)^{3/2}\exp(-\frac{\Delta_\m}{\kB T}),
    \end{equation}
where we have re-expressed the monopole mass in terms of the monopole bandwidth $W_\m=\hbar^2/(ma^2)$. 
The real part of the conductivity in Eq.~\eqref{sigmaomega} is  
    \begin{equation}\label{eq:ReSigma}
        \frac{\Re[\sigma(\omega)]}{\sigma_0}=
        \sigma_\text{peak}
        \frac{\Gamma^2\omega^2 }%
        {(\omega^2-\omega_p^2)^2+\Gamma^2 \omega^2},
    \end{equation} 
where 
    \begin{align}
        \omega_p^2(T)&= 
        \frac{n_\m(T)(Q_\m^2\mu^{-1}a)}{m}
        \approx
        \frac{2Q_\m^2}{(2\pi)^{3/2}}\,\frac{\mu^{-1}W_\m}{\hbar^2}
        \left(\frac{\kB T}{W_\m}\right)^{3/2}\exp(-\frac{\Delta_\m}{\kB T}),
    \\
        \Gamma(T)&=
        % \frac{4\pi\,Q_\m^4}{135}
        % \left(\frac{\mu^{-1}}{mc^2}\right)^2
        % \frac{\hbar}{ma^2}
        % \left(\frac{\kB T}{W_\phi}\right)^4
        % =
        \frac{4\pi\,Q_\m^4}{135}
        \left(\frac{W_\m\mu^{-1}}{W_\phi^2}\right)^2
        \left(\frac{\kB T}{W_\phi}\right)^4
        \frac{W_\m}{\hbar},
    \\
        \text{and }
        \sigma_\text{peak}(T)&=
        \frac{2\pi W_\m}{\hbar\Gamma(T)}\cdot a^3n_\m(T)
        \approx
        \frac{135}{
        Q_\m^4 (2\pi)^{3/2}
        }
        \left(\frac{\mu^{-1}}{W_\phi}\right)^{-2}
        \left(\frac{W_\m}{W_\phi}\right)^{-7/2}
        \left(\frac{\kB T}{W_\phi}\right)^{-5/2}\exp(-\frac{\Delta_\m}{\kB T})
    \end{align}
Defining a rescaled frequency $x=\omega/\omega_p$, we can rewrite Eq.~\eqref{eq:ReSigma} as
    \begin{equation}\label{eq:sigmax}
        \frac{\Re[\sigma(\omega)]}{\sigma_0}
        =
        \sigma_\text{peak}
        \frac{x^2}
        {\mathcal{Q}^2(x^2-1)^2+x^2}
    \end{equation}
where we have introduced the ``quality factor'' $\mathcal{Q}$, defined as
    \begin{equation}
        \mathcal{Q}(T)\coloneq
        \frac{\omega_p(T)}{\Gamma(T)}
        =
        \frac{135}{\sqrt{2}(2\pi)^{7/4}\,Q_\m^3}
        \left(\frac{\mu^{-1}}{W_\phi}\right)^{-3/2}
        \left(\frac{W_\m}{W_\phi}\right)^{-13/4}
        \left(\frac{\kB T}{W_\phi}\right)^{-13/4}\exp(-\frac{\Delta_\m}{2\kB T})
    \end{equation}
%Note that we do not assign any interpretation to $\mathcal{Q}$ as a ratio of stored to dissipated energy as in usual oscillator systems; it serves as a convenient parameter that controls the line-shape of $\sigma(\omega)$.  
%%
\begin{figure}[t]
    \centering
    \includegraphics[width=0.5\columnwidth]{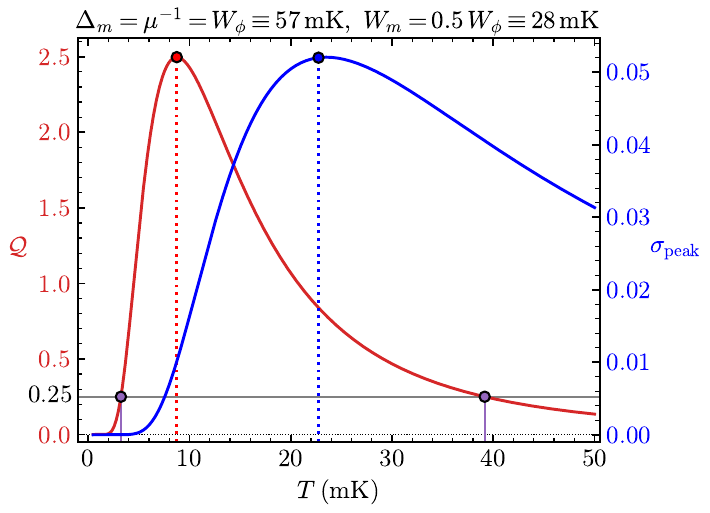}
    \caption{Quality factor $\mathcal{Q}$ (in red, left axis) and peak conductivity $\sigma_\text{peak}$ (in blue, right axis) as a function of temperature in the resonance regime ($\mu^{-1}=\Delta_\m=W_\phi\equiv57\text{mK}$, $W_\m\equiv28\text{mK}$). Both traces are non-monotonic; $\mathcal{Q}$ is maximal at $T\approx 0.15\Delta_\m\equiv8.7\text{mK}$ while $\sigma_\text{peak}$ is maximal at $T\approx 0.4\Delta_\m\equiv22.7\text{mK}$.}
    \label{fig:Qsigma}
\end{figure}

Interestingly, both the peak conductivity $\sigma_\text{peak}$ and the quality factor $\mathcal{Q}$ are non-monotonic with temperature, as shown in Fig.~\ref{fig:Qsigma}. 
Defining a rescaled variable $\tau=\kB T/\Delta_\m$, either quantity scales as $\tau^{-s}\exp[-1/(b\tau)]$ for appropriate choices of $s$ and $b$. The latter function has a maxima at $\tau=1/(sb)$, which implies that $\sigma_\text{peak}(T)$ is maximal at $\kB T=2/5\Delta_\m$ whereas $\mathcal{Q}$ is maximal at $\kB T=2/13\Delta_\m$. 
Furthermore, both $\mathcal{Q}$ and $\sigma_\text{peak}$ are strongly influenced by the monopole bandwidth $W_\m$; they increase as $W_\m$ narrows.

If we posit that the resonance peaks are visible when $\mathcal{Q}>\mathcal{Q}_\text{threshold}$, then such a criterion is satisfied only in a temperature window, $[T_\text{low},T_\text{high}]$, as a consequence of the non-monotonicity. 
Above $T_\text{high}$ the quality factor degrades due to increased dissipation, whereas below $T_\text{low}$ the exponentially suppressed monopole density reduces $\mathcal{Q}$. 
In the case where all the monopole energy-scales in the problem are comparable to the photon bandwidth, the maximal quality factor turns out to be approximately $0.25$. We choose $\mathcal{Q}_\text{threshold}=0.25$ and indicate $T_\text{low}$ and $T_\text{high}$ in Fig.~\ref{fig:Qsigma}.

\begin{figure}[!t]
    \centering
    \includegraphics[width=\columnwidth]{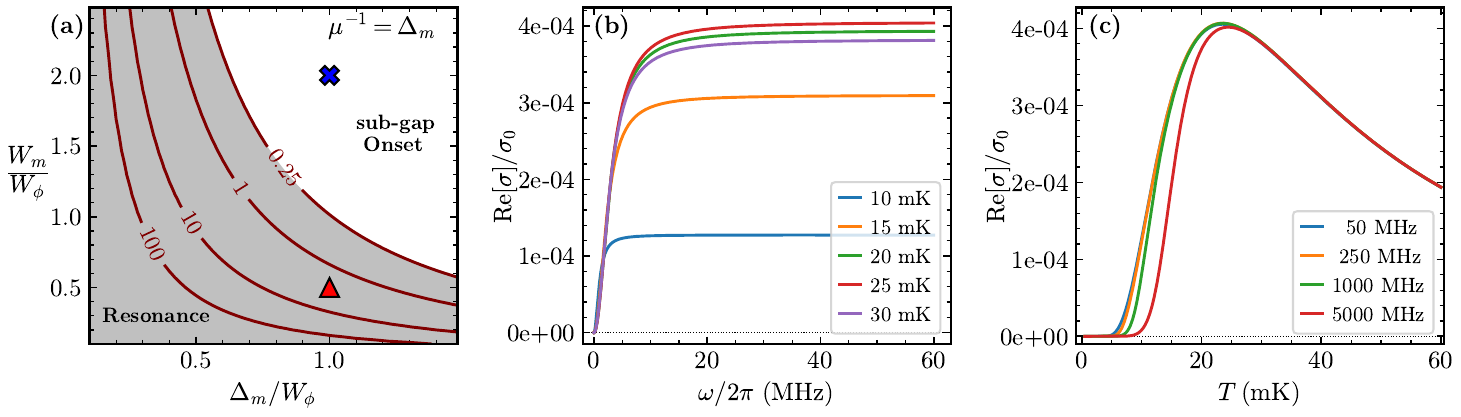}
    \caption{\textbf{(a)} Contour plot of maximal quality factor $\mathcal{Q}_\text{max}$ as a function of monopole gap $\Delta_\m$ and bandwidth $W_\m$. The resonance region (shaded in gray) is separated from the sub-gap onset regime by the $\mathcal{Q}_\text{max}=0.25$ contour. Other level curves (in dark red) show contours at fixed $\mathcal{Q}_\text{max}=1,10$, and $100$. 
    The red triangle at $(\Delta_\m,W_\m)=(W_\phi,0.5 W_\phi)$ is a representative point of the resonance regime; conductivity plots with these parameters are reported in the main text. 
    The blue cross at $(\Delta_\m,W_\m)=(W_\phi,2 W_\phi)$ is a representative point in the onset regime. 
    For this choice of parameters we plot the real part of conductivity \textbf{(b)} as a function of frequency and \textbf{(c)} as a function of temperature.
    }
    \label{fig:onset}
\end{figure}

We will show that the expression in Eq\eqref{eq:sigmax} predicts two qualitatively different regimes for conductivity, as delineated in Fig.~\ref{fig:onset}(a). 
The function $f(x)=x^2/[\mathcal{Q}^2(x^2-1)^2+x^2]$ in Eq.~\eqref{eq:sigmax} always reaches its maximum at $x=1$, i.e. $\sigma(\omega)$ has a maximum at $\omega=\omega_p$. 
It is straightforward to establish that the function reaches half its maximum at $x_\pm=\sqrt{1+\frac{1}{4\mathcal{Q}^2}}\pm\frac{1}{2\mathcal{Q}}$. 
When $\mathcal{Q}\gg1$, $x_\pm\approx 1\pm\tfrac{1}{2\mathcal{Q}}$, which signifies a resonant peak with $\text{FWHM}=\omega_p(x_+-x_-)\approx \Gamma$. 
However, in the opposite limit when $\mathcal{Q}\ll1$, $x_-\approx \mathcal{Q}$, and $x_+\approx 1/\mathcal{Q}$, i.e. the function rapidly reaches half its maximum at a low frequency $\omega=\omega_p\mathcal{Q}\ll \omega_p$, and remains almost plateau-like until a high frequency $\omega\approx\omega_p/\mathcal{Q}=\Gamma$. We call this the onset regime. 

To summarize, the conductivity has two regimes :
(i) \textbf{Resonance :} 
for low dissipation ($\mathcal{Q}\gtrsim 1$) the conductivity has a low frequency peak at $\omega=\omega_p$, and 
(ii) \textbf{sub-gap Onset :} for high dissipation ($\mathcal{Q}\ll 1$) a broad shoulder like feature sets in at $\omega\sim \omega_p^2/\Gamma$ as depicted in Fig.~\ref{fig:onset}(b). 
At temperatures much below the monopole gap $\Delta_m$, both of these features occur at very low frequency ($10$-$100$s of MHz) compared to the insulating charge gap in the system.

\end{document}